\documentclass[sigconf]{acmart}

\AtBeginDocument{%
  }

\usepackage{booktabs}
\usepackage{bm}
\usepackage{tikz}
\usetikzlibrary{calc}
\usepackage{amsmath}
\usepackage{lipsum}
\usepackage{array}
\usepackage{color}
\usepackage{graphicx}
\usepackage{diagbox}
\usepackage{balance}
\usepackage{adjustbox}
\usepackage{tikz}
\usepackage{pgfplots}
\usetikzlibrary{patterns}
\usetikzlibrary{spy,backgrounds}
\usepgfplotslibrary{groupplots}
\usepackage{pgfplotstable}
\pgfplotsset{compat=1.12}
\usepackage{subfigure}
\usepackage{multirow}
\usepgfplotslibrary{statistics}
\usepackage{ctable}
\usepackage{booktabs,caption}
\usepackage[flushleft]{threeparttable}
\usepackage[ruled,linesnumbered]{algorithm2e}
\usepackage{epstopdf}
\usepackage{algorithmicx}
\usepackage{algpseudocode}
\usepackage{multicol}
\usepgfplotslibrary{units}
\usepackage{fancybox}
\usepackage{hyperref}
\usepackage{pgfplotstable}
\usepackage{color}
\usepackage{wasysym}
\usepackage{pifont}
\usepackage{listings}
\usepackage{balance}

\usepackage{tabularx}
\usepackage{hyperref}
\hypersetup{
colorlinks=true,
linkcolor=blue,
citecolor = blue,      
urlcolor=cyan,
breaklinks=true,
}

\SetKwInput{KwInput}{Input}                
\SetKwInput{KwOutput}{Output}

\definecolor{RYB1}{RGB}{044, 045, 084}
\definecolor{RYB2}{RGB}{067, 068, 117}
\definecolor{RYB3}{RGB}{107, 108, 163}
\definecolor{RYB4}{RGB}{150, 155, 199}
\definecolor{RYB5}{RGB}{135, 188, 189}
\definecolor{RYB6}{RGB}{137, 171, 124}
\definecolor{RYB7}{RGB}{111, 153, 084}
\definecolor{RYB8}{RGB}{154, 205, 50}
\definecolor{RYB9}{RGB}{169, 169, 169}

\newcommand{\leakagePattern}[1]{\ensuremath{\mathsf{#1}}}
\newcommand{\tvol}{\leakagePattern{tvol}\xspace}
\newcommand{\fvol}{\leakagePattern{fvol}\xspace}
\newcommand{\rlen}{\leakagePattern{rlen}\xspace}
\newcommand{\ulen}{\leakagePattern{ulen}\xspace}
\newcommand{\ilen}{\leakagePattern{ilen}\xspace}
\newcommand{\dlen}{\leakagePattern{dlen}\xspace}
\newcommand{\qeq}{\leakagePattern{qeq}\xspace}
\newcommand{\rid}{\leakagePattern{rid}\xspace}
\newcommand{\co}{\leakagePattern{co}\xspace}
\newcommand{\qcor}{\leakagePattern{qcor}\xspace}

\SetCommentSty{mycommfont}

\newcommand{\lei}[1]{{\textcolor{blue} {#1}}}
\newcommand{\rev}[1]{{\textcolor{black} {#1}}}

\newcommand{\revs}[1]{{\textcolor{black} {#1}}}

\copyrightyear{2023}
\acmYear{2023}
\setcopyright{acmlicensed}\acmConference[CCS '23]{Proceedings of the 2023 ACM SIGSAC Conference on Computer and Communications Security}{November 26--30, 2023}{Copenhagen, Denmark}
\acmBooktitle{Proceedings of the 2023 ACM SIGSAC Conference on Computer and Communications Security (CCS '23), November 26--30, 2023, Copenhagen, Denmark}
\acmPrice{15.00}
\acmDOI{10.1145/3576915.3623085}
\acmISBN{979-8-4007-0050-7/23/11}

\begin{document}

\title{Leakage-Abuse Attacks Against Forward and Backward Private Searchable Symmetric Encryption}

\author{Lei Xu}
\orcid{0000-0001-9178-6640}
\affiliation{%
  \institution{
  Nanjing University of Science and Technology}
  \city{Nanjing}
  \country{China}
}
\affiliation{%
  \institution{City University of Hong Kong}
  \city{Hong Kong}
  \country{China}
}
\email{xuleicrypto@gmail.com}

\author{Leqian Zheng}
\orcid{0000-0003-3772-4400}
\affiliation{%
  \institution{
  City University of Hong Kong}
  \city{Hong Kong}
  \country{China}
}
\email{leqizheng2-c@my.cityu.edu.hk}

\author{Chengzhi Xu}
\orcid{0009-0007-9102-9464}
\affiliation{%
  \institution{
  Nanjing University of Science and Technology}
  \city{Nanjing}
  \country{China}
}
\email{xcz@njust.edu.cn}

\author{Xingliang Yuan}
\orcid{0000-0002-3701-4946}
\affiliation{%
  \institution{
  Monash University}
  \city{Melbourne}
  \country{Australia}
}
\email{xingliang.yuan@monash.edu}

\author{Cong Wang}
\orcid{0000-0003-0547-315X}
\affiliation{%
  \institution{
  City University of Hong Kong}
  \city{Hong Kong}
  \country{China}
}
\email{congwang@cityu.edu.hk}

\renewcommand{\shortauthors}{Lei Xu, Leqian Zheng, Chengzhi Xu, Xingliang Yuan, \& Cong Wang}

\begin{abstract}
Dynamic searchable symmetric encryption (DSSE) enables a server to efficiently search and update over encrypted files. To minimize the leakage during updates, a  security notion named forward and backward privacy is expected for newly proposed DSSE schemes. Those schemes are generally constructed in a way to break the linkability across search and update queries to a given keyword. However, it remains underexplored whether forward and backward private DSSE is resilient against practical leakage-abuse attacks (LAAs), where an attacker attempts to recover query keywords from the leakage passively collected during queries.

In this paper, we aim to be the first to answer this question firmly through two non-trivial efforts. First, we revisit the spectrum of forward and backward private DSSE schemes over the past few years, and unveil some inherent constructional limitations in most schemes. Those limitations allow attackers to exploit query equality and establish a guaranteed linkage among different (refreshed) query tokens surjective to a candidate keyword. Second, we refine volumetric leakage profiles of updates and queries by associating each with a specific operation. By further exploiting update volume and query response volume, we demonstrate that all forward and backward private DSSE schemes can leak the same volumetric information (e.g., insertion volume, deletion volume) as those without such security guarantees. To testify our findings, we realize two generic LAAs, i.e., frequency matching attack and volumetric inference attack, and we evaluate them over various experimental settings in the dynamic context. Finally, we call for new efficient schemes to protect query equality and volumetric information across search and update queries. 
\end{abstract}





\maketitle

\section{Introduction}
\noindent Dynamic searchable symmetric encryption (DSSE)~\cite{KamaraPR12,KamaraP13,CashJJJKRS14,abs-2201-05006} enables a server to efficiently search and update over encrypted files. During queries and updates (file addition and deletion), the server only knows precisely defined leakage, but not the contents of queries and files. 
In the past decade, various DSSE schemes have actively been proposed with regard to improved functionality~\cite{PappasKVKMCGKB14,DemertzisPPDG16,KamaraM17,LaiPSLMSSLZ18}, security ~\cite{NaveedPG14,WangZ16,StefanovPS14}, and efficiency~\cite{KamaraP13,CashJJJKRS14,abs-2201-05006,Bossuat21}. 
Among others, minimizing the leakage for DSSE is the most emerging objective in the literature. 
A security notion of DSSE named forward and backward privacy (FP/BP)~\cite{StefanovPS14, BostMO17} has become essential and standard for newly designed DSSE schemes~\cite{BostMO17,SunYLSSVN18,ChamaniPPJ18,SunSLYSLNG21,ChamaniPKD22,Amjad23}. It echos the intrinsic security goals of SSE, i.e., no information about data should be learned if it has not been searched.  

At a high level, forward privacy implies that newly added files cannot be associated with queries in the past, and the goal is to protect previous queries and files during updates to them; while backward privacy implies that subsequent queries cannot be executed over already deleted files, and the goal is to protect files and updates to them during queries. 
To achieve those security notions, all existing forward and backward private DSSE schemes are constructed in a way to break the linkability across queries and updates to a given keyword. As a result, leakage in those schemes is expected to only reveal minimally necessary information in updates and queries, such as the number of updates and the number of matched results for a query.

Because of the above security features, a common understanding today is that FP/BP-DSSE may possibly mitigate file injection attacks~\cite{ZhangKP16,PoddarWLP20,BlackstoneKM19,ZhangWPLK23}. 
In these attacks (aka chosen-data attacks), an active attacker can inject carefully chosen files into the encrypted dataset, causing each keyword to manifest unique patterns that could potentially lead to the recovery of encrypted query keywords.
%
However, beyond file injection attacks - a specific type of leakage-abuse attacks (LAAs)~\cite{CashGPR15,KamaraKMSTY22}, the study on those modern DSSE schemes and their resilience against more generic passive LAAs remains largely underexplored.

Leakage-abuse attacks (LAAs) against searchable symmetric encryption (SSE) schemes~\cite{IslamKK12, CashGPR15, BlackstoneKM19, XuDZYW21, NingHPYL0D21, KamaraKMSTY22, Gui21PP23} show the implication and impact of leakage in real-world adversarial settings, where an attacker could have access to the background information of the underlying dataset. 
%
%
By formalizing the relationship between such background knowledge and the adversarial view using matrices or bi-partite graphs, it is demonstrated that the fundamental reason why LAAs can succeed is: the underlying characteristics of a dataset remain intact after the encryption of SSE~\cite{BlackstoneKM19}. Throughout queries, the attacker can identify these unique and structurally-invariant characteristics (e.g., query response lengths, query co-occurrence counts) that exist both in unencrypted and encrypted datasets, so as to facilitate query recovery.

Implicitly, \rev{for any LAA to be effective}, the attackers have to determine which collected leakage patterns (from query tokens and query responses) are mapped to the underlying keywords. \rev{Here, the query token refers to the encrypted representation of the keyword to be queried, acting as the credential for the search operation.} \revs{The primary purpose of using tokens in SSE is to enable the server to conduct keyword searches on encrypted data stored without disclosing the underlying plaintexts.} We note that all passive LAAs investigate static SSE, where updates are excluded. 
In the static setting, the mapping between the set of query tokens and their respective responses, and the set of candidate keywords in the plaintext database are bijective. Namely, each query token and its response are exactly associated with a keyword. This makes the query recovery equivalent to finding the best bijective mapping between those two sets, using the aforementioned structural-equivalent leakage and the attacker's background knowledge. 

\begin{figure}
  \includegraphics[width=\linewidth]{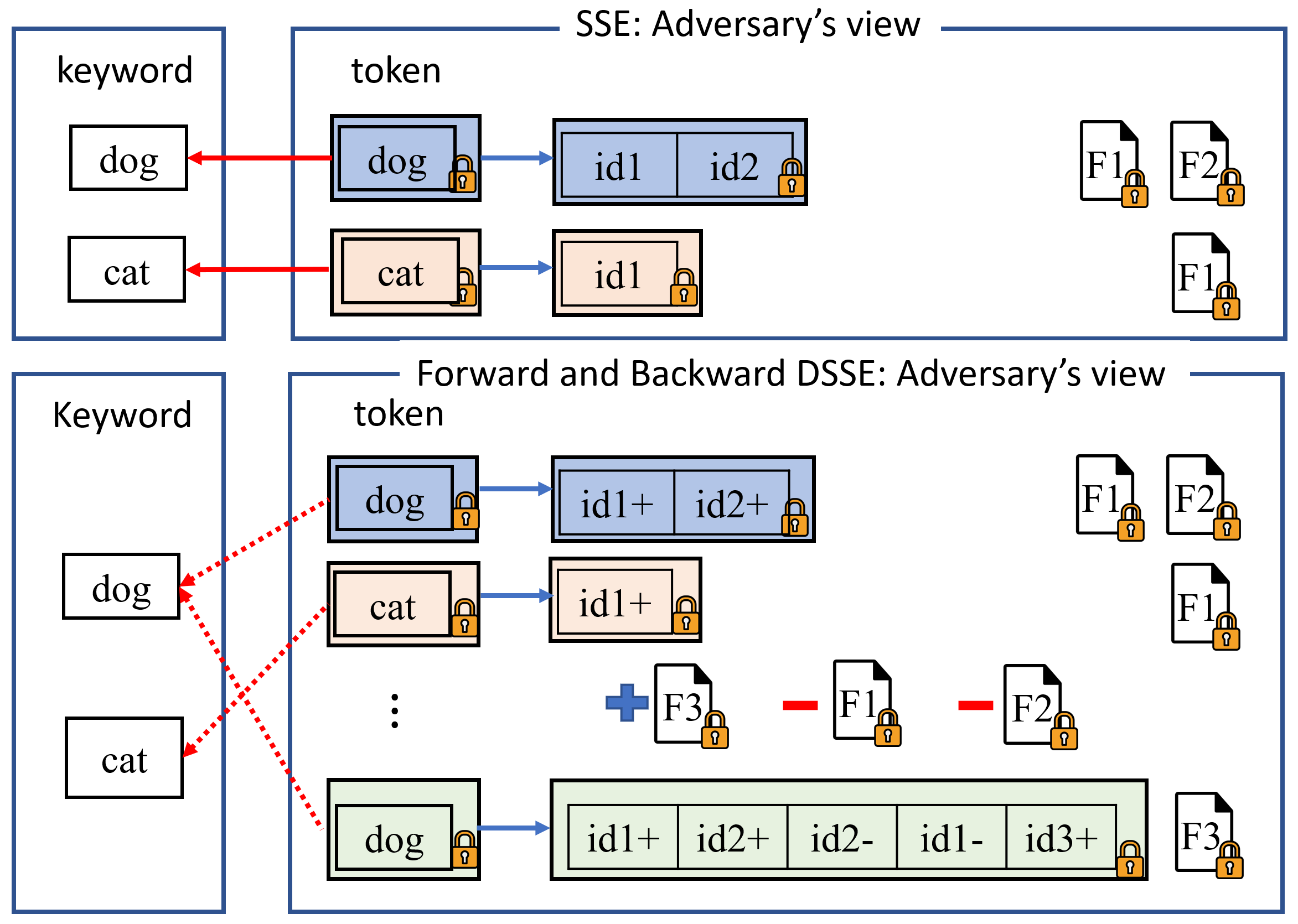}
  \caption{Difference between static SSE and forward and backward private DSSE}\label{fig:dynamic}
  \vspace{-10pt}
\end{figure}

In dynamic settings, this is no longer the case due to refreshed query tokens and responses, which are demanded by forward and backward security notions. As seen in Fig.~\ref{fig:dynamic}, there exist multiple individual query tokens and their respective responses related to the same candidate keyword. 
That is, the mapping from the set of observed query tokens and their responses to the set of candidate keywords becomes surjective but no longer bijective. 
Moreover, the client may add and/or remove files without the control of an attacker. That naturally obfuscates the exposed leakage from queries and updates and further increases the difficulty of passive LAAs.

\begin{table*}[!ht]
\begin{threeparttable}
\caption{\rev{A comparison of existing attacks and our attacks against (D)SSE schemes}}\label{tab:attack overview}
\begin{tabularx}{\textwidth}{*{3}{l} *{2}{c}}
\toprule
\multicolumn{2}{c}{Attack Setting\tnote{1}} & {Attack}  & {Leakage Exploited\tnote{2}} & {Attack Target\tnote{3}} \\
\midrule
\multirow{10}[3]{*}{Passive against SSE} & \multirow{6}{*}{Known-data}  & Count attack~\cite{CashGPR15} & \co, \rlen  &  Na\"ive  \\
& & ${\rm Subgraph^{ID}}$~\cite{BlackstoneKM19} & \rid &  Na\"ive  \\
& & ${\rm Subgraph^{VL}}$~\cite{BlackstoneKM19} & \fvol  & Na\"ive  \\
& & VolAn~\cite{BlackstoneKM19} & \tvol & Na\"ive  \\
& & SelVolAn~\cite{BlackstoneKM19} & \fvol, \rlen  & Na\"ive \\
& & LEAP~\cite{NingHPYL0D21} & \co & Na\"ive  \\
\cmidrule[0.001em]{2-5}
& \multirow{4}{*}{Sampled-data} & IKK~\cite{IslamKK12} & \co   & Na\"ive \\
& & SAP~\cite{Simon21} & \rlen, \qeq & CLRZ~\cite{ChenLRZ18}, SEAL~\cite{DemertzisPPS20}, PPYY~\cite{PatelPYY19}  \\
& & IHOP~\cite{OyaK22} & \qcor, \co, \qeq & CLRZ~\cite{ChenLRZ18}, Pancake~ \cite{GrubbsKLBL0R20}, OSSE~\cite{ShangOPK21}  \\
& & Gui et al.~\cite{Gui21PP23} & \co & CLRZ~\cite{ChenLRZ18}, PPYY~\cite{PatelPYY19}, PRT-EMM\cite{KamaraM19} \\
\midrule[0.001em]
\multirow{4}{*}{Active against DSSE} & \multirow{4}{*}{Chosen-data} & Decoding\&Binary~\cite{BlackstoneKM19} & \tvol & Na\"ive   \\
& & \citet{ZhangKP16} & \rid & Na\"ive \\
& & BVA~\cite{ZhangWPLK23} & \fvol & SEAL~\cite{DemertzisPPS20}, ShieldDB~\cite{VoYSLNW23}  \\
& & BVMA~\cite{ZhangWPLK23} & \rlen, \qeq & SEAL\cite{DemertzisPPS20}, ShieldDB~\cite{VoYSLNW23}  \\
\midrule[0.003em]
\multirow{3}{*}{Passive against DSSE} & Sampled-data & $\mathsf{FMA}$ & \qeq, \fvol & FP/BP-DSSE \cite{BostMO17,ChamaniPPJ18,SunYLSSVN18,SunSLYSLNG21,ChamaniPKD22}, VH/FP/BP-EMM~\cite{Amjad23} \\  
& Known-data & $\mathsf{VIA}$ and $\mathsf{PVIA}$  & \fvol, \rlen & FP/BP-DSSE \cite{BostMO17,ChamaniPPJ18,SunYLSSVN18,SunSLYSLNG21,ChamaniPKD22}, VH/FP/BP-EMM~\cite{Amjad23} \\ 
& Sampled-data &$\mathsf{LVIA}$  & \rlen  & FP/BP-DSSE \cite{BostMO17,ChamaniPPJ18,SunYLSSVN18,SunSLYSLNG21,ChamaniPKD22} \\ 
\bottomrule
\end{tabularx}
\begin{tablenotes}
\item[1] Chosen-data attacks require the adversary to choose some of the data in the data collection. Known-data attacks require knowledge of a subset of the data collection. Sampled-data attacks require a statistical distribution of the data collection, such as keyword distribution of the data collection or the frequency of the data collection to be accessed;
\item[2] \co, \rid, \fvol, \tvol, \rlen, \qcor, and \qeq denote the co-occurrence pattern, response identifier pattern, file volume pattern, total volume pattern, response length pattern, query correlation pattern, and query equality pattern, respectively;
\item[3] It lists the constructions that the attack targets. Na\"ive, FB/BP and VH refer to SSE constructions that leak full access pattern, achieve forward privacy/backward privacy, and response length hiding, respectively. Here, VH denotes the response length (the number of files matching the query) hiding.  
\end{tablenotes}
\end{threeparttable}
\end{table*}

\subsection{Our Contributions}

\revs{Given the above observations and challenges, in this paper, we aim to first answer the following open question: Is forward and backward dynamic SSE more resilient against LAAs, compared to static SSE? 
%
To answer this question, we thoroughly revisit the spectrum of existing forward and backward private DSSE schemes over the past few years~\cite{BostMO17,SunYLSSVN18,ChamaniPPJ18,SunSLYSLNG21,ChamaniPKD22,Amjad23}. Our conclusion presents a seemingly counterintuitive perspective to the above question: \emph{those DSSE schemes with advanced security properties are also vulnerable to LAAs to a similar extent, compared to the traditional static schemes.}}

Our key observations lay from two aspects. 
First, there are inherent constructional limitations in specific schemes that allow attackers to establish guaranteed linkage between refreshed query tokens after encrypted updates and their historical counterparts. 
Second, more generally, the leakage during updates, especially the volume of updates, if not protected well, will allow attackers to combine it with other leakage patterns (some are refined and formalized in this work) to group those refreshed query tokens and their responses that are more likely to be associated with a candidate keyword. 
In other words, both observations enable the attacker to effectively group seemingly unrelated query tokens (and their respective query responses) together. The grouping of query tokens transforms the aforementioned surjective mapping into a bijective one between the set of query tokens (and responses) and the set of candidate keywords. Thus, the attacker can readily launch known LAAs using the structurally-invariant leakage against forward and backward private DSSE, just like static SSE. 

Below, we elaborate on our technical contributions. 

\noindent{\bf Refining Leakage Profiles for DSSE.}  
The first is how to identify and exploit the structurally-invariant leakage in the dynamic setting.  
Learning from prior LAAs, we reckon that the core information for exploitation expects to be the volumetric information of queries and updates. 
We note that existing LAAs targeting static SSE exploit singular leakage profiles, which are insufficient for DSSE. 
As opposed to merely using a response volume pattern, we refine and formalize the volumetric information further by associating it with specific operations. 
In particular, volumetric leakage patterns are defined as \emph{insert volume pattern, delete volume pattern, response volume pattern, and update volume pattern}. The update volume pattern refers to that of both addition and deletion operations. 

It is worth noting that forward and backward private DSSE schemes do not separately leak insert volume and delete volume in updates, as operations are hidden during updates. 
But by exploiting query equality, update volume, and response volume, we find that those two leakage profiles can trivially be derived, as demonstrated in Section~\ref{subsec:vol int}.
It turns out that all forward and backward private DSSE schemes leak the same volumetric information as those without such security guarantees.
We also show that those refined leakage profiles can help identify clients' updates, differentiating whether the client executes deletion operations or not. 
Inspired by very recent work~\cite{Gui21PP23}, we further consider a system-wide leakage, i.e., the file volume pattern, which is inherent in any SSE schemes during file retrieval. 
We find that one can exploit this leakage to predict whether or not two different query tokens are for the same keyword, even if they originally cannot be linked from a specific scheme, e.g., ORAM-like constructions.

\noindent{\bf Revisiting Forward and Backward Private DSSE Schemes.}
With the above build leakage framework, we thoroughly revisit forward and backward private DSSE schemes~\cite{BostMO17,SunYLSSVN18,ChamaniPPJ18,SunSLYSLNG21,ChamaniPKD22,Amjad23} in the literature. 
As mentioned, we discover that several schemes allow one to link refreshed query tokens of the same keyword from the specific construction level. 
In some forward private constructions, it is known that a query token can normally be computed by that of the latter one. 
In backward privacy, schemes can be classified into three levels, from the strongest to the weakest, according to the security they can provide. 
In particular, except for schemes with the strongest security level (mostly ORAM-like constructions), the equality of queries can directly or indirectly be leaked, e.g., query tokens with respect to the same keyword share some common components. More details can be found later in Section~\ref{subsec:qeq}.
Once query equality can be determined, connections between queries and updates to a given keyword can also be derived. 

\noindent{\bf Realizing LAAs against Forward and Backward Private DSSE.} Based on the above stepping stones, we realize two generic passive query recovery attacks, frequency matching attack (FMA) and volumetric inference attack (VIA), under different settings and assumptions. 
%
%
In particular, FMA assumes that only the query distribution information is known~\cite{Simon21,Damie0P21}. Unlike prior work, FMA first builds query equality information from observed queries and then determines the frequencies of \rev{linked queries that have the same underlying keyword.} After that, the attacker can match the collected query frequencies with the known query distribution for query recovery. It is noteworthy that the attack performance of FMA is intrinsically augmented by the heterogeneous query distributions across distinct temporal intervals in dynamic settings. To be specific, when the query distribution pertains solely to a singular time interval, the corresponding recovery rate is a mere 1.8\%. In contrast, when the query distributions are provided across 24 temporal intervals, the enhanced recovery rate spans a range of 63.5\% to 93.4\% across diverse settings. %

VIA adapts to common assumptions of prior LAAs~\cite{CashGPR15,BlackstoneKM19,NingHPYL0D21}, and considers attackers with different levels of background knowledge, i.e., a fraction of files in updates between two linked queries, or the distribution of the dataset.
After deriving query equality, VIA exploits our refined leakage profiles to recover the observed queries. 
Experimental results indicate that the recovery rate can reach up to $36.8\%$, even when only $50\%$ of the inserted files are accessible to the attacker and up to $30\%$ of the files are removed by the client. Furthermore, the recovery rate can increase up to $96\%$ when more background knowledge is provided to the attacker. \rev{Table~\ref{tab:attack overview} overviews the recent attacks against SSE schemes. As summarized, our work is the first to focus on passive LAAs against DSSE schemes with forward and backward privacy.}

\noindent{\bf Takeaway Messages.} This work uncovers often-overlooked leakage patterns in existing forward and backward private dynamic SSE schemes. 
By incorporating dynamic operations into the traditional volumetric leakage and query equality leakage patterns, we refine the previous leakage patterns and bridge the gap between static SSE and dynamic SSE schemes. 
With this new leakage framework, we demonstrate how to recover queries in passive settings, illustrating that forward and backward privacy are insufficient and vulnerable to leakage-abuse attacks. Our research highlights the need for improved security measures against the proposed LAAs, urging the development of efficient schemes to safeguard query equality and volumetric information during search and update queries.

\section{Related Work}\label{sec:related work}
\noindent{\bf Dynamic Searchable Symmetric Encryption.} As mentioned above, DSSE is motivated to support fundamental database operations like addition and deletion. Like static SSE, since the first DSSE constructions given by Kamara et al.~\cite{KamaraPR12}, research interests on DSSE still focuses on asking for better trade-offs between the efficiency with lower search complexity and client storage overhead~\cite{KamaraP13,CashJJJKRS14,abs-2201-05006,Bossuat21,KamaraMPQ21}, stronger security property with less leakage~\cite{NaveedPG14,WangZ16,StefanovPS14,VoLYSNL20,ChamaniPKD22} and expressiveness of \rev{supported queries}~\cite{PappasKVKMCGKB14,DemertzisPPDG16,KamaraM17,LaiPSLMSSLZ18}. Despite the numerous constructions being proposed, including those that support forward and backward privacy, it has been pointed out by Patel et al.~\cite{PatelPY20} that they either rely on the use of oblivious RAM with at least logarithmic cost~\cite{GargMP16} or tolerate specific leakages to obtain efficiency improvement~\cite{KamaraMO18,GeorgeKM21}. 

\noindent{\bf Leakage Abuse Attack and Suppressions.} Leakage abuse attack (LAA) is a widely concerned threat that explores the real-world security problem of SSE in deployment. 
The main idea of LAA is to leverage the structurally-invariant characteristics that can be observed by the adversary to match their background knowledge to recover the query or reconstruct the clear dataset. 
The first seminar work on LAA was proposed by Islam et al.~\cite{IslamKK12} and later improved by~\citet{CashGPR15}. After that, a line of works under different adversarial models are proposed to attack SSE and other property-preserving secure search schemes~\cite{KellarisKNO16,NaveedKW15,DurakDC16,PouliotW16,GrubbsSB0R17,LachariteMP18,BindschaedlerGC18,GrubbsLMP18,GuiJW19,KornaropoulosPT19,GrubbsLMP19,0015YWWX19,FalzonMACRST20,PanELWQM0A20,KornaropoulosPT20,MarkatouFTS21,NingHPYL0D21,KornaropoulosPT21,Simon21,OyaK22}. 
To mitigate LAA, the primary task is to suppress the above leakages~\cite{WangSLQ022,PatelPYY19,DemertzisPPS20,GrubbsKLBL0R20,0019ZD00J22,XuDZYW21}, and the notion of forward and backward privacy was born in this context~\cite{BostMO17,SunYLSSVN18,SunSLYSLNG21,ChamaniPPJ18,ChamaniPKD22}. 
Recent years have witnessed increasing efforts on forward and backward private DSSE with the hope of being efficient~\cite{SunSLYSLNG21,ChamaniPPJ18}. 
And most of these follow-up works initiate the studies with the same security notion like~\cite {BostMO17}. 
In addition to these works, there are also other works attempting to find more secure BP-DSSE schemes or define stronger backward security notions. 
For example,~\citet{ChamaniPPJ18}'s construction $\mathsf{Orion}$ does not reveal the total number of updates associated with the given keyword. 
The state-of-the-art, volume hiding, forward and backward private DSSE schemes were introduced by~\citet{Amjad23}. They tolerate query equality patterns to obtain efficiency improvements.

\section{Preliminaries}\label{sec:preliminaries}
\noindent We present a background of dynamic searchable encryption and its corresponding security notions.

\noindent{\bf Basic Notations.}
Let $\mathcal{W}$ be a keyword space and ${\bf D}$ be a file collection. Let $\mathsf{DB} = \{(\mathsf{id}_i, \mathsf{W}_i)\}$ be a set of records that describe the correlations between keywords and files, where $\mathsf{W}_i$ denotes the set of all keywords appearing in ${\rm D}_i \in {\bf D}$ and $\mathsf{id}_i$ represents the identifier of ${\rm D}_i$. We define $\mathsf{DB}(w)$ as the set of files containing $w$, i.e., $\mathsf{DB}(w) = \{\mathsf{id}_i: w \in \mathsf{W}_i\}$. Lastly, we introduce the intersection of two collections. Given two collections $\mathsf{X} = \{x_1, \dots, x_m\}$ and $\mathsf{Y} = \{y_1, \dots, y_n\}$, we define their intersection $\mathsf{Z}$ as the collection of elements, including duplicities, appearing \rev{in both collections}. For example, given $\mathsf{X} = \{1, 1, 2, 2\}$ and $\mathsf{Y} = \{1, 2, 2\}$, then their intersection is $\mathsf{Z} = \mathsf{X} \Cap \mathsf{Y} = \{1, 2, 2\}$. Likewise, their union is defined as $\mathsf{W} = \mathsf{X} \Cup \mathsf{Y} = \{1,1,2,2\}$. For ease of presentation, we refer to the number of items in a set as $|\cdot|$ and the number of items in a collection as $\#(\cdot)$. Clearly, we have $\#({\rm D}_i) \ge |\mathsf{W}_i|$.

\noindent{\bf Dynamic Searchable Symmetric Encryption.} A dynamic searchable symmetric encryption scheme $\mathsf{DSSE} $ $= (\mathsf{Setup}$, $\mathsf{Search, Update})$ is a tuple of one algorithm and two protocols between the client and the server proceeding as follows:

\begin{itemize}
 \item  $(\mathsf{K}, \sigma, \mathsf{EDB}) \leftarrow \mathsf{Setup}(\mathsf{DB})$ is a probabilistic polynomial-time algorithm that takes $\mathsf{DB}$ as inputs and outputs a secret key $\mathsf{K}$ and a secret state $\sigma$ for the client and an encrypted database $\mathsf{EDB}$ for the server. 
 \item $(\mathsf{R}, \bot) \leftarrow \mathsf{Search}(\mathsf{K}, \sigma,  q, \mathsf{EDB})$ is a protocol run between the client with the key $\mathsf{K}$, its state $\sigma$, and a query $q$ as inputs, and the server with $\mathsf{EDB}$ as inputs. 
 At the end of the protocol, the client receives a set of files $\mathsf{R}$ and the server receives nothing. 
 In this paper, we only consider single-keyword search schemes where $q$ refers to a keyword $w$.
 \item $(\sigma^{\prime}, \mathsf{EDB}^{\prime})\leftarrow \mathsf{Update}(\mathsf{K}, \sigma, \mathsf{op}, \mathsf{in}, \mathsf{EDB})$ is protocol run between the client with the key $\mathsf{K}$, its state $\sigma$, an update operation $\mathsf{op}$, and an update data $\mathsf{in}$ as inputs,
 and the server with the encrypted database $\mathsf{EDB}$ as input. 
 At the end of the protocol, the client will update its state as $\sigma^{\prime}$, and the server will update its encrypted database as $\mathsf{EDB}^{\prime}$. 
 The update operation $\mathsf{op}$ is taken from $\{\mathsf{add}, \mathsf{del}\}$ denoting the addition and the deletion of a keyword-identifier pair, respectively. 
\end{itemize}
A $\mathsf{DSSE}$ scheme is \emph{perfectly correct} if the search protocol returns all files matching the query. Notice that we focus on the DSSE schemes with perfect correctness in this work. 


\noindent{\bf Security.} The security of DSSE defines what an adversary learns in a DSSE scheme. It can be formalized as the indistinguishable model between a real word game $\mathsf{SSE}_{\mathsf{Real}}$ and an ideal world game $\mathsf{SSE}_\mathsf{Ideal}$ with predefined leakage profiles. These leakage profiles are usually modeled as triple $\mathcal{L} = (\mathcal{L}_{\mathsf{Setp}}, \mathcal{L}_{\mathsf{Srch}}, \mathcal{L}_{\mathsf{Updt}})$, defining what information from $\mathsf{Setup}, \mathsf{Search}, \mathsf{Update}$ leaks to the adversary. A DSSE scheme requires that the adversary learns nothing about the issued query and the content of the encrypted database except the above-admitted leakage profiles. The formal security notion of the DSSE scheme is presented in Appendix~\ref{appendix:sec-notion}.

\noindent {\bf Forward and Backward Private DSSE.} Despite the above security notion built on indistinguishability, DSSE also requires forward and backward privacy due to the disclosed update pattern and access pattern. 

For forward privacy, it requires that the newly added file/keyword pairs cannot be linked by previous query tokens. This implies that previous query tokens cannot be used to search the above newly added entries. In existing constructions, such a notion is usually achieved by refreshing the query token. While for backward privacy, it focuses on the privacy of entries which are added and deleted later. \rev{In consideration of the balance between security and efficiency, Bost et al.~\cite{BostF17} classify backward privacy into three levels, ranked from the strongest to the weakest. The details of these levels are as follows:}
\begin{itemize}
    \item Type-I. Backward privacy with insertion pattern: reveals the files that currently match keyword $w$, along with the time of their insertion and the total number of updates.
    \item Type-II. Backward privacy with update pattern: leaks the files currently matching keyword $w$, the time of their insertion, and the time of all updates made on $w$ (excluding their content) are revealed.
    \item Type-III. Weak backward privacy: discloses the files that currently match keyword $w$, along with the time of their insertion, the time of all updates made on $w$, and which deletion update corresponds to which insertion update.
\end{itemize}
 
The formal notions of forward privacy and three levels of backward privacy are given in Appendix~\ref{appendix:forward and backward}.

\section{Refining Leakage Profiles}
\noindent In this section, we will refine the volumetric leakage profiles further to accommodate both insert and delete operations in dynamic settings. The primary purpose of making this refinement is to analyze what forward and backward private DSSE leaks from a volumetric perspective, which has not drawn sufficient attention.

\subsection{Modeling Leakage}\label{subsec:modleak}
\noindent Given a $\mathsf{DB}$ and a \rev{query list $\mathcal{Q}$ in the form of $(\mu, w)$}, we first present leakage patterns that have been defined in~\cite{BostF17}, \rev{where $\mu$ is the timestamp and $w$ denotes the search keyword.} Specifically,
\begin{itemize}
\item $\mathsf{TimeDB}(w)$ denotes the list of all files that match the query $q$, excluding deleted ones, together with the timestamp of when each file was inserted. Formally,
\begin{align*}
\mathsf{TimeDB}(w)=\{(\mu, \mathsf{id})| (\mu, \mathsf{add}, (w, \mathsf{id})) \in \mathcal{Q}\\~~~~~~~~~{\rm and}~\forall \mu^{\prime}, (\mu^{\prime}, \mathsf{del}, (w, \mathsf{id}))\notin \mathcal{Q}\}.
\end{align*}
\item $\mathsf{Updates}(w)$ denotes the list of the timestamps of updates on $w$. Formally,
\begin{align*}
    \rev{\mathsf{Updates}(w)} = \{\mu| (\mu, \mathsf{add}, (w, \mathsf{id}))~\\~~~~~~{\rm or}~(\mu, \mathsf{del}, (w, \mathsf{id}))\in\mathcal{Q}\}.
\end{align*}
\item $\mathsf{DelHist}$ denotes the list of timestamps for all deletion operations and the timestamp of the inserted entry it removes. Formally,
\begin{align*}
    \mathsf{DelHist}(w) = \{(\mu^{\mathsf{add}}, \mu^{\mathsf{del}})| \exists~\mathsf{id}~{\rm s.t.}~(\mu^{\mathsf{del}}, {\mathsf{del}}, (w, \mathsf{id})) \\ \in \mathcal{Q}~{\rm and}~ (\mu^{\mathsf{add}}, \mathsf{add}, (w, \mathsf{id})) \in \mathcal{Q}\}.
\end{align*}
\end{itemize}

With the update operations and their corresponding timestamp in mind, next, we put forward some new types of leakage that provide a more detailed description of volumetric information in dynamic settings. The following are the details:
\begin{itemize}
\item \rev{{\bf Update length pattern} denoted as $\mathsf{ulen}(w) = |\mathsf{Updates}(w)|$ outputs the number of updates (i.e., $\mathsf{add}$ and $\mathsf{del}$) made on keyword $w$. }
\item \rev{{\bf Insert length pattern} denoted as $\mathsf{ilen}(w) = |\mathsf{Updates}(w)|-|\mathsf{DelHist}(w)|$ outputs the number of insertions (i.e., $\mathsf{op} = \mathsf{add}$) made on $w$.}
\item \rev{{\bf Delete length pattern} denoted as $\mathsf{dlen}(w) = |\mathsf{DelHist}(w)|$ outputs the number of deletions (i.e., $\mathsf{op} =\mathsf{del}$) made on $w$.}
\item \rev{{\bf Response length pattern} (aka result length pattern) denoted as $\mathsf{rlen}(w)=|\mathsf{TimeDB}(w)|$ outputs the number of files matching keyword $w$.}
\item {\bf File volume pattern} denoted as $\mathsf{fvol}(w) = \{\#{\rm D}_{\mathsf{id}}| \mathsf{id} \in $ $ \mathsf{TimeDB}(w) \}$ outputs the volume of each file containing $w$. Usually, we measure the volume by the number of keywords each file contains.
\item \rev{{\bf Response similarity pattern} denoted as $\mathsf{rsp}(w_i, w_j)$ outputs the similarity between two query responses. Here, we measure the similarity by the extended Jaccard similarity coefficient for collections, i.e., $\mathsf{rsp}(w_i, w_j)=\#(\mathsf{fvol}(w_i) \Cap \mathsf{fvol}(w_j))/\#(\mathsf{fvol}(w_i)\Cup\mathsf{fvol}(w_j))$.}
\item \rev{{\bf Query equality pattern} (aka {\rm search pattern}) denoted as $\mathsf{qeq}(w_i, w_j) = \bm{1}(w_i=w_j)$ indicates whether two queries are targeting the same keyword.
$\bm{1}(\cdot)$ is the indicator function outputting 1 if the input evaluates to true and 0 otherwise.}
\end{itemize}
As seen above, we first link the volumetric information with specific update operations and incurring timestamps and define the insertion length patterns, update length patterns and delete length patterns. Besides, we also introduce a new notion termed response similarity pattern, which is used to quantify the similarity of the query responses. Later, we will show that these refined leakage patterns will play a critical role in evaluating the security of the forward and backward DSSE schemes.

\subsection{Volumetric Leakages in BP-DSSE}\label{subsec:vol int}
\noindent The above section recalls the definition and the difference between the three-level of backward security in terms of access patterns and timestamps. In this section, we will revisit the above three tuples from the volumetric aspect and interpret them from a new perspective. The reason \rev{why we are doing this} is that a line of volume attacks have been proposed in the last few years and they are shown to cause devastating damage to existing searchable encryption schemes. And we find that these damages are getting worse in dynamic scenarios, because the volume information changes over time (as seen in Figure~\ref{fig:length-change}), and the adversary can use therein difference to help recover the query further.

Combined with the definition listed in Appendix~\ref{appendix:forward and backward}, we now show the volumetric leakage in different levels of backward private notions. Here we pay our attention to four types of volumetric information, which are defined as follows:
\begin{itemize}
  \item $n_w$ - the number of files matching $w$
  \item $u_w$ - the total number of updates on $w$
  \item $a_w$ - the number of insertions on $w$
  \item $d_w$ - the number of deletions on $w$.
\end{itemize}
Observe that since the total number of updates is the sum of insertions and deletions, and the response length is the difference between insertions and deletions, thus we have the following equations, i.e., 
\begin{equation}\label{equ:num}
\begin{cases}
n_w = a_w - d_w \\
u_w = a_w + d_w 
\end{cases} 
\end{equation}
Then if one has any two of these four values, the equation can be transformed into a system of two binary linear equations. Solving this equation system can obtain the other two values. In all three levels of backward privacy notions, we have observed that each of them leaks two of the aforementioned values in their search procedures. Specifically, 
\begin{itemize}
\item Type-I backward private constructions disclose volumetric information $u_w$ and $n_w = |\mathsf{TimeDB}(w)|$,
\item Type-II backward private constructions reveal volumetric information $n_w=|\mathsf{TimeDB}(w)|$ and $u_w=|\mathsf{Updates}(w)|$, 
\item Type-III backward private constructions leak volumetric information $n_w = |\mathsf{TimeDB}(w)|$ and $d_w=|\mathsf{DelHist}(w)|$.
\end{itemize}
This implies that all three-level backward privacy notions share the same volumetric leakage. For example, for the Type-I backward private constructions revealing $n_w$ and $u_w$, we can obtain 
\begin{equation}\label{equ:numAdvanced}
\begin{cases}
n_w = a_w - d_w \\
u_w = a_w + d_w 
\end{cases} \Longrightarrow
\begin{cases}
a_w=(u_w+n_w)/2 \\
d_w=(u_w-n_w)/2 
\end{cases}.
\end{equation}
Similarly, we can get all four types of volumetric information for the other two levels of backward private construction. To this end, once an attacker possesses one of the above four types of background knowledge, it can use the corresponding leakage to infer the information of the query. It could be noted that the ways to derive the above leakages differ in different constructions. Following is an example to show how to derive the above volumetric information in a classical backward private DSSE construction proposed by~\citet{BostF17}.

\begin{figure}
  \includegraphics[width=\linewidth]{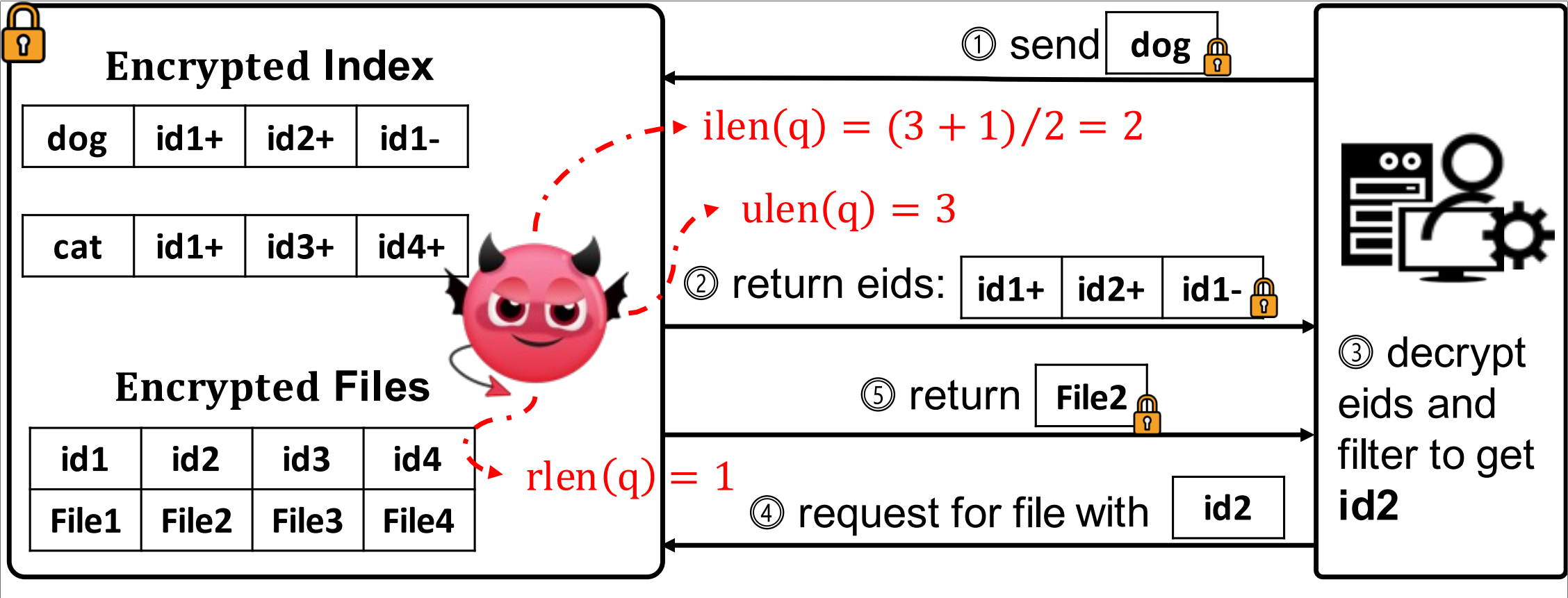}
  \caption{Example of extracting volumetric leakages from $\mathsf{B}({\Sigma})$ or $\mathsf{B}(\mathsf{O})$~\cite{BostMO17}}\label{fig:Round-DSSE}
  \vspace{-10pt}
\end{figure}

\noindent{\bf Example: } To clearly present these leakages, \rev{we take Bost et al.'s generic two-roundtrip solution~\cite{BostMO17} as an example to show how to extract volumetric leakage from the constructions.} It should be noted that their solutions can provide at least Type II BP and will achieve Type-I backward privacy if ORAM or TWORAM techniques are adopted. For the purposes of presentation, we refer to the generic solution as $\mathsf{B}({\Sigma})$ and the ORAM-based solution as $\mathsf{B}(\mathsf{O})$. Since the ORAM also does not protect volumetric information, we choose the former as an example. We first review the main idea of it. The diagram in Figure~\ref{fig:Round-DSSE} shows the index structure and query operation in $\mathsf{B}({\Sigma})$. The client builds the encrypted index for the tuple $(w, \mathsf{id}\|\mathsf{op})$ from an arbitrary SSE scheme. Note that, here a label $\mathsf{op}$ (aka ``$\mathsf{+}$'', ``$\mathsf{-}$'' in the figure) is attached to the file identifier to remark the operation to the tuple is addition or deletion.    

Combined with Figure~\ref{fig:Round-DSSE}, we now show how to obtain the above leakages. In the query stage, one can retrieve the encrypted files containing the given keyword by a two-roundtrip interaction. Firstly, the client uses the generated token to fetch the encrypted indices that match the query with the help of the server. In this round, the total number of updates on the keyword \ulen$=3$ can be observed. Once these indices are obtained, the client executes the decrypt operation locally and filters out those which have not been deleted before. Then it sends these undeleted indices to the server, and the server returns the corresponding encrypted files to the client. This procedure discloses the actual number of files in the results, i.e.,  \rlen$=1$. According to Equation~(\ref{equ:numAdvanced}), it is easy to know that the real number of files containing the given keyword injected before is 2.

\subsection{Query Equality Pattern}\label{subsec:qeq}
\noindent Another leakage pattern we want to revisit is the query equality pattern, which is the key to computing the query frequency pattern. In previous works, one can always obtain the query equality pattern by checking whether they keep the same search token. While in the context of forward and backward private DSSE, it does not work because the search token in most schemes will be refreshed upon the update operations. Nevertheless, this does not mean that forward and backward private DSSE will not leak query equality patterns. As pointed out by Kammar~\cite{GeorgeKM21}, if an adversary observes that some query $q_1$ has the largest volume $v$, then observes an edit operation, and observes that some query $q_2$ now has volume $v + 1$, it can deduce that $\mathsf{qeq}(q_1, q_2)=1$ and the query equality is leaked. The following parts provide more systematic thoughts on rebuilding the above query equality pattern in forward and backward private DSSE schemes. 

Generally, there is no ``one-size-fits-all'' approach to determine the $\mathsf{qeq}$ due to the different instantiation principles of FB-DSSE schemes and therein query status. By surveying existing constructions, we find that one can rebuild the query equality pattern through these two aspects: 1) \textit{the relations of tokens}; 2) \textit{the similarity of responses}. 

For the former one, throughout recent forward and backward construction, we find that the relations of tokens with respect to the same keyword can be classified into the following four categories:
\begin{itemize}
\item \emph{exactly same} means that the tokens with respect to the same queried keyword are the same.
\item \emph{partially same} means that tokens with respect to the same keyword share some same components.
\item \emph{computationally same} means that one token can be computed by that of the later one.
\item \emph{irrelevant} indicates that no correlation can be found between the two queries.
\end{itemize}
With these relations, one can determine $\mathsf{qeq}$ of the queries by checking whether their tokens satisfy one of the above first three conditions. If yes, we have $\mathsf{qeq} =1$, and 0 otherwise. 

Table~\ref{tab:token relations} summarizes \rev{the leakage patterns revealed by selected FP/BP-DSSE constructions and how they expose the query equality pattern}. As seen, $\mathsf{B}({\Sigma})$~\cite{Bost16} uses the same token for the same keyword and encrypts them to prevent the server from learning indices. The plaintext indices can only be obtained and processed by the client. Thus, the server cannot get any sensitive information about updates without the client. A typical example of partially same tokens is \citet{SunYLSSVN18}'s scheme, whose tokens with respect to the same keywords are partially same. Tokens in this scheme are in the form of $\mathsf{T}_w=(F(K,w), msk)$, where $F(K,w)$ is the same with respect to the same keyword. They keep the freshness of tokens by updating $msk$ after each query to guarantee forward privacy.

\begin{table}[t]\small
\begin{threeparttable}
\caption{\rev{Leakage of representative FP/BP-DSSE schemes}}\label{tab:token relations}
\centering
\begin{tabularx}{\linewidth}{l|cccccc|r}
\toprule
\multirow{2}[1]{*}{Scheme}   & \multicolumn{6}{c|}{\rev{Leakage Pattern}\tnote{1}} & \multirow{2}[1]{*}{Security} \\ \cmidrule{2-7} & \leakagePattern{qeq}\tnote{2}  & \rlen & \ulen & \ilen & \dlen & \fvol &  \\
\midrule
$\mathsf{B}(\mathsf{O})$~\cite{BostMO17}\tnote{3}   &     $\overset{?}{\approx}$  & \CIRCLE  & \CIRCLE & \CIRCLE & \CIRCLE & \CIRCLE  & \multirow{3}{*} {Type~\uppercase\expandafter{\romannumeral1}} \\
$\mathsf{Moneta}$~\cite{BostMO17}    &    $\overset{?}{\approx}$   &  \CIRCLE & \CIRCLE & \CIRCLE & \CIRCLE & \CIRCLE &\\
$\mathsf{Orion}$~\cite{ChamaniPPJ18} &     $\overset{?}{\approx}$   & \CIRCLE  & \Circle & \Circle & \Circle & \CIRCLE & \\
\midrule[0.001em]
$\mathsf{B}({\Sigma})$~\cite{BostMO17}\tnote{4}     & $\equiv$    &  \CIRCLE & \CIRCLE & \CIRCLE & \CIRCLE & \CIRCLE & \multirow{5}{*}{Type~\uppercase\expandafter{\romannumeral2}}\\
$\mathsf{Fides}$~\cite{BostMO17}     &   $\overset{\mathcal{C}}{=}$   & \CIRCLE & \CIRCLE & \CIRCLE & \CIRCLE & \CIRCLE & \\ 
$\mathsf{Aura}$~\cite{SunSLYSLNG21}  &    $\overset{\mathcal{P}}{=}$   & \CIRCLE & \CIRCLE & \CIRCLE & \CIRCLE & \CIRCLE & \\
$\mathsf{Mitra}$~\cite{ChamaniPPJ18} &   $\overset{\mathcal{P}}{=}$    & \CIRCLE & \CIRCLE & \CIRCLE & \CIRCLE & \CIRCLE & \\
$2\mathsf{ch}_{\bf FB}$~\cite{Amjad23}      &   $\overset{\mathcal{P}}{=}$  & \Circle & \Circle & \Circle & \Circle & \CIRCLE &   \\
\midrule[0.001em]
$\mathsf{Janus}$~\cite{BostMO17}     &   $\overset{\mathcal{P}}{=}$  & \CIRCLE &  \CIRCLE & \CIRCLE & \CIRCLE & \CIRCLE & \multirow{4}{*}{Type~\uppercase\expandafter{\romannumeral3}}\\
$\mathsf{Janus}$++~\cite{SunYLSSVN18} &    $\overset{\mathcal{P}}{=}$  & \CIRCLE & \CIRCLE & \CIRCLE & \CIRCLE & \CIRCLE & \\
$\mathsf{OSSE}$~\cite{ChamaniPKD22}   &   $\overset{\mathcal{P}}{=}$  & \CIRCLE & \CIRCLE & \CIRCLE & \CIRCLE & \CIRCLE & \\
$\mathsf{LLSE}$~\cite{ChamaniPKD22}   &    $\overset{\mathcal{P}}{=}$  & \CIRCLE & \CIRCLE & \CIRCLE & \CIRCLE & \CIRCLE & \\
\bottomrule
  \end{tabularx}
\begin{tablenotes}
\item[1] The listed leakage patterns are defined when these constructions are used in file-retrieval applications, where $\CIRCLE$ denotes that the listed scheme exposes the corresponding leakage pattern, and $\Circle$ signifies that it does not;
\item[2] $\overset{?}{\approx}$ denotes that tokens are irrelevant but we can infer \qeq with their response similarity. $\equiv, \overset{\mathcal{C}}{=}, \overset{\mathcal{P}}{=} $ refer to exactly same, computationally same, and  partially same, respectively, which describes the relations of tokens for the same keyword in listed constructions;
\item[3] The generic backward-private scheme initiated with ORAM; 
\item[4] The generic two-roundtrip backward-private scheme initiated with an arbitrary SSE scheme.
\end{tablenotes}
  \end{threeparttable}
  \vspace{-10pt}
\end{table}

Another way we give through query response is proposed for schemes (like $\mathsf{Moneta}$~\cite{BostMO17}) that do not leak the relations of the tokens. In this case, the client always deletes the indices searched before and rebuilds a new one. To deal with this, we usually leverage the similarity between the responses to build the equality pattern. For robustness, here we introduce another way of using file volume patterns to compute the response similarity. Specifically, for any two queries $q$ and $q^{\prime}$ with file volume leakage $\mathsf{fvol}_q$ and $\mathsf{fvol}_{q^{\prime}}$, we denote their response similarity as
\begin{equation}
    {\mathsf{rsp}}({q, q^{\prime}}) = {\#(\mathsf{fvol}_q \Cap \mathsf{fvol}_{q^{\prime}})}/{{\#(\mathsf{fvol}_q \Cup\mathsf{fvol}_{q^{\prime}})}}.
\end{equation}
Based on the fact that queries for the same information usually have similar responses, here we suggest estimating the query equality pattern by using a threshold method. Specifically, we define
\begin{equation}
\mathsf{qeq} (q, q^{\prime}) = \begin{cases}
1,\quad &{\mathsf{rsp}}({q, q^{\prime}}) \geq \delta \\
0,\quad &{\mathsf{rsp}}({q, q^{\prime}}) < \delta
\end{cases}
\end{equation}
where the threshold is characterized by $\delta$, \rev{which can be inferred from the growth rate of data volume. Specifically, we set $\delta = \min \{|\mathsf{DB}_{\mu_i}|/|\mathsf{DB}_{\mu_j}|, |\mathsf{DB}_{\mu_j}|/|\mathsf{DB}_{\mu_i}|\}$ in this work, where $|\mathsf{DB}_{\mu_i}|$ denotes the size of dataset at time $\mu_i$.}  In reality, the volume of updated data is relatively small compared to the entire database, thus we can set $\delta > 0.5$ in this work. The same setting will be adopted in our experiment.

\subsection{Attack Model and Overview}
\noindent \rev{In Section~\ref{sec:related work}}, we have shown that the potential risk brought by leakage to static SSE has been confirmed by many works. Does this still hold for dynamic SSE (with forward and backward security)? 
%
%
Before giving a formal answer, we first overview the attack model.

\noindent{\bf Targets and Attack Settings.} Differing from the previous attacks on DSSE schemes, we focus on {\bf passive attacks}. The attacker captures certain metadata (also known as auxiliary information) of encrypted datasets, such as data distribution, query frequency, etc. He is allowed to monitor the transcripts between the client and the server and exploits his background knowledge (e.g., distribution, statistics) about the updates. Like most of the existing LAAs, our goal is to recover the observed queries by the disclosed leakage.

In practice, it is rational to consider the attacker can access the background knowledge of updates. \rev{For instance, in the context of the COVID-19 pandemic, statistics like daily increases in the number of new cases can be obtained from social media. This allows attackers to gain volumetric information related to updates, even if the underlying database of medical records is encrypted. 
Geospatial datasets could also serve as an example, as the location information is often publicly available. For instance, the US Places of Worship dataset records all places of worship across states. The data classifies meters by the restriction type and includes location descriptions. Additionally, if one adopts SSE techniques to build encrypted databases, then the transaction log of the underlying database engine would reveal information such as when and how encrypted files are updated w.r.t. which (deterministically) encrypted keywords. There is indeed work~\cite{GuiK23USENIX} about how to exploit transaction logs against SSEs, while the attacking target is not forward/backward private.}

{\noindent}{\bf Attack Models.} Based on different attack settings, auxiliary data needed, and leakage profile exploited, LAAs can be classified into different categories. In this paper, we focus on the following two common types of attacks:

\begin{itemize}
  \item {\bf Frequency Matching Attack ($\mathsf{FMA}$).} Given a series of queries and query distribution in the period that encompasses all target queries, $\mathsf{FMA}$ utilizes the equality pattern to recover the query information. The key to this attack is to utilize exposed leakages to calculate the query equality pattern as well as the query frequency.
  \item {\bf Volumetric Inference Attack ($\mathsf{VIA}$).} Given a series of queries, $\mathsf{VIA}$ exploits volumetric leakage to recover the query. To perform the attack, the attacker must possess some prior knowledge of the encrypted dataset, such as a fraction of plaintext data collection or its distribution. This prior knowledge can be the dataset before the time $\mu_0$ or the updated information between two \rev{linked queries that have the same underlying keyword.}
\end{itemize}
From the target of the attack, both two attacks discussed are commonly referred to as query recovery attacks. 
In the subsequent sections, we will comprehensively validate our observations mentioned of forward and backward DSSEs by presenting the above-mentioned two types of attacks. We will thoroughly present the scope and details of each attack in the following sections.

\section{Frequency Matching Attack}\label{sec:FMA}
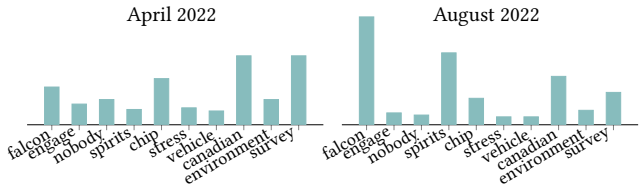
\begin{figure}[bt]
\center
\subfigure{
\hspace{-10pt}\begin{tikzpicture}[scale=0.48]
\begin{axis}[ybar,
    legend style={ legend columns=-1,
    legend pos=north east
    },  
    xtick=data,
    bar width=4mm,
    ymin=0,
    width=0.55\textwidth,
    height=0.3\textwidth,
    xticklabels from table={\wordtable}{name},
    x tick label style={rotate=30,anchor=east,font=\huge},
    tick label style={font=\huge},
    y axis line style={opacity=0},
    axis x line*=bottom,
    legend style={font=\large},
    label style={font=\large},
    xlabel style={yshift=-5ex},
    ymax=0.4,
    hide y axis,
    area legend]    
    \addplot [RYB5,fill=RYB5,x tick label style={xshift=-0.3cm}] table[x=Clusters,y=Length] {\aprtable};
    \end{axis}
    \draw (4,3)  node[font=\small]{April 2022};
\end{tikzpicture}
}
\hspace{-10pt}\subfigure{
\begin{tikzpicture}[scale=0.48]
\begin{axis}[ybar,
    legend style={ legend columns=-1,
    legend pos=north east
    },  
    xtick=data,
    bar width=4mm,
    ymin=0,
    width=0.55\textwidth,
    height=0.3\textwidth,
    xticklabels from table={\wordtable}{name},
    x tick label style={rotate=30,anchor=east,font=\huge},
    tick label style={font=\huge},
    y axis line style={opacity=0},
    axis x line*=bottom,
    legend style={font=\large},
    label style={font=\large},
    xlabel style={yshift=-5ex},
    hide y axis,
    ymax=0.4,
    area legend]    
    \addplot [RYB5,fill=RYB5,x tick label style={xshift=-0.3cm}] table[x=Clusters,y=Length] {\augtable};
    \end{axis}
        \draw (4,3)  node[font=\small]{August 2022};
\end{tikzpicture}
}
\vspace{-20pt}
 \caption{Query distribution of 10 randomly selected keywords in April and August 2022}\label{fig:period-fre}
\vspace{-15pt}
\end{figure}

\noindent We start our work with the $\mathsf{FMA}$ attack, which matches the collected query frequencies with the known query distribution for query recovery. 
The key to this attack is to build the query equality for observed queries, and then the frequency of each query performed can be calculated. Intuitively, the success of $\mathsf{FMA}$ significantly hinges on the diversity of the query frequency. Affected by factors such as seasonality, trending topics, news events, day of the week, and month of the year~\cite{Googletrends111}, the query distribution usually changes over time, providing necessary diversity (as seen in Figure~\ref{fig:period-fre}). This diversity can be leveraged to identify the query further by comparing the reconstruction spaces and computing their intersection in multiple rounds. More details are offered below. 

\noindent{\bf Attack Assumption.} Let ${\bm q}=\{q_1, \dots, q_t\}$ be observed queries and ${\mathcal D}=$ $\{{\bm d}_{[t_i,t_{i+1}]} \sim (p^{(i)}_{w_1}, \cdots, p^{(i)}_{w_{|\mathcal{W}|}})\}_{i=1}^{\tau}$ be distribution of queries over $\mathcal{W}$ in different time intervals. In $\mathsf{FMA}$,  we consider a passive and persistent attacker who is additionally aware of the auxiliary information of the keyword frequency requested. \rev{The goal of $\mathsf{FMA}$ is to recover the true keywords that generated the query sequence ${\bm q}=\{q_1, \dots, q_t\}$.}

\pgfplotstableread{
Clusters Length 
1        0.111357751 
2        0.061043285  
3        0.07436182 
4       0.045135035 
5        0.136145024 
6        0.049944506  
7       0.040695523  
8       0.203477617  
9   0.07436182
10      0.203477617    
}\aprtable


\pgfplotstableread{
Clusters Length 
1 0.318336163
2 0.035016978
3 0.028650255
4 0.212224109
5 0.078098472
6 0.023344652
7 0.023344652
8 0.142826825
9 0.042657046
10 0.095500849
}\augtable

\pgfplotstableread{
name
falcon
engage
nobody
spirits
chip
stress
vehicle
canadian
environment
survey
}\wordtable

\noindent{\bf Attack Description.} As mentioned, the fundamental principle of $\mathsf{FMA}$ includes identifying the query equality of these queries, calculating query frequency, and comparing them with the provided distribution to recover queries. 

In accordance with this principle, our first step is to leverage the approach provided in Section~\ref{subsec:qeq} to build the query equality for queries within the same time interval (\lei{lines 19-25}). Once the query equality pattern is built, we put queries with the query equality pattern of 1 in the same group. For example, given query sequences $\tilde{\bm q}$ with query distribution $\tilde{\bm d}\sim(\tilde{p}^{(i)}_{w_1}, \cdots, \tilde{p}^{(i)}_{w_{|\mathcal{W}|}})$ in a certain time interval, if $\mathsf{qeq}(\tilde{q}_i, \tilde{q}_j)=1$, $\tilde{q}_i$ and $\tilde{q}_j$ will be grouped together. Once the groups, denoted by $\mathcal{G}=\{G\}$, are generated, our second step is to build the candidate set for each group (\lei{lines 26-28}).  In fact, the query frequency of the keyword issuing in a group is the size of that group. According to this, keywords whose requested frequencies are closest to the size of $G$ will be assigned as the candidates for queries in $G$. Namely, for $\tilde{q} \in G$, we have
\begin{equation*}
  \mathsf{MM}[q] = \arg\min\limits_{w \in \mathcal{W}}{|\#(G)-\#(\tilde{\bm q})\cdot \tilde{p}_w|}
\end{equation*}
where $\mathsf{MM}$ is the multi-map from the query to its candidates.

By applying the above procedure to query sequences at different time intervals, a candidate set can be generated for each query (\lei{lines 1-9}). In this process, the query keyword can be determined if only one element remains in the set (\lei{lines 5-8}). For queries that do not meet this condition, it continues to narrow down candidates across different time intervals. Specifically, we select queries related to the same keyword in different time intervals based on their query equality and compute the intersection of their candidate set (\lei{line 11}). If the candidate set remains only one element, it is the output as the information for the query. The pseudo-code of $\mathsf{FMA}$ is detailed in the Algorithm~\ref{Alg:FMA}.

Note that, $\mathsf{FMA}$ is effective to forward and backward DSSE schemes that hide responses or re-encrypt indices after the query. This is because refreshing the query response does not alter file volume information, thus preserving the query equality information.

\begin{algorithm}[t]\small
\DontPrintSemicolon 
  \caption{\textbf{Frequency Matching Attack}}\label{Alg:FMA}
  \KwInput{Encrypted database $\mathsf{EDB}$, keyword space $\mathcal{W}$, a query list ${\bm q}=\{q_1, \dots, q_t\}$, and query distribution in different time intervals ${\mathcal D}=$ $\{{\bm d}_{[t_i,t_{i+1}]} \sim (p^{(i)}_{w_1}, \cdots, p^{(i)}_{w_{|\mathcal{W}|}})\}_{i=1}^{\tau}$}
  \KwOutput{A map $\mathsf{M}$ from ${\bm q}$ to $w \in \mathcal{W}$.}
  Initialize an empty map $\mathsf{M}$ and an empty multi-map $\mathsf{MM}^*$ \; 
  Partition ${\bm q}$ to ${\bm q}^{(1)}$, $\cdots, {\bm q}^{(\tau)}$ according to $\{t_i, t_{i+1}\}_{i \in [\tau]}$ \;
  \For{{\rm query sequences} ${\bm q}^{(i)}$ {\rm incur in} $[t_i, t_{i+1}]$}{
  Run $\mathsf{MM}^{(i)} \leftarrow \mathsf{Candidate\_Gen}(f, {\bm q}^{(i)})$\;
  \If{$\mathsf{MM}^{(i)}[{q}]= 1$}{
  Set $\mathsf{M}[q] \leftarrow \mathsf{MM}^{(i)}[q]$ and remove $q$ from ${\bm q}$\;
  Remove $\mathsf{MM}^{(i)}[q]$ from $\mathsf{MM}^{(i)}$
  }
  }
  \For{{\rm remained} $q \in {\bm q}$}{
  Set $\mathsf{MM}^*[q] = \bigcap_{q^{\prime} \in {\bm q}} \mathsf{MM}^{(i)}[q^{\prime}]$ where $\mathsf{qeq}(q, q^{\prime})=1$\;
  \If{$|\mathsf{MM}^*[q]|=1$}{
  Add all $(q^{\prime}, \mathsf{MM}^*[q])$ to $\mathsf{M}^*$ where $\mathsf{qeq}(q, q^{\prime})=1$
  }
  }
  \KwRet query-keyword maps $\mathsf{M}$\;
  \SetKwFunction{FQue}{$\mathsf{Candidate\_Gen}$}
  \SetKwProg{Fn}{Function}{:}{}
  \Fn{\FQue{\rm{$\tilde{\bm d}$}, $\tilde{\bm q}$}}{
  Initialize an empty multi-map $\mathsf{MM}$ and an initial cluster $\mathcal{G} = \{G_1 = \{\tilde{q}_1\}\}$\;
  \For{{\rm each} $\tilde{q}_i \in \tilde{\bm q}$}{
  \If{$\exists~\tilde{q}_j \in G_k, G_k \in \mathcal{G}$ {\rm s.t.} $\mathsf{qeq}(\mathsf{EDB}, \tilde{q}_i, \tilde{q}_j)=1$}{
  Add $\tilde{q}_i$ into $G_k$
  \Else{
  Create a new group $G_{|\mathcal{G}+1|} = \{\tilde{q}_j\}$ 
  }
  }
  }
  \For{{\rm each} $\tilde{q} \in G \in \mathcal{G}$}{
    Set $\mathsf{MM}[\tilde{q}] \leftarrow \arg \min_{w \in \mathcal{W}}|\#(G)- \#(\tilde{\bm q})\tilde{p}_w|$\;
  }
  \KwRet query-keyword multi-maps $\mathsf{MM}$
  }
\end{algorithm}

\section{Volumetric Inference Attack}\label{sec:tiv}
\noindent The second attack is the volumetric inference attack ($\mathsf{VIA}$) which exploits volumetric information to recover the query. 
For attack practicality, we will consider attackers with different levels of knowledge.

\subsection{$\mathsf{VIA}$: Volumetric Inference Attack}
We initiate the attack from a baseline assumption analog to that of existing LAAs where the attacker has knowledge of the added files between two linked queries \rev{that have
the same underlying keyword}~\cite{CashGPR15,BlackstoneKM19}. Studying the vulnerability of $\mathsf{VIA}$ under this assumption is of theoretical significance in evaluating the strength of the proposed scheme against a powerful adversary. We then gradually constrain this assumption to design more practical attacks.

\noindent{\bf Attack Assumption.} Let $q$ denote a query performed on an encrypted database $\mathsf{EDB}$ that the attacker is targeting. Let $\tilde{q}$ be a query requesting the same keyword after $q$. According to the analysis in Section~\ref{subsec:qeq}, such a $\tilde{q}$ can always be identified if it exists. Let $\overline{\mathsf{D}}$ be the file collections inserted into $\mathsf{EDB}$ between $q$ and $\tilde{q}$, which is known to the attacker. Suppose a sequence of operations is performed on $\overline{\mathsf{D}}$ \rev{by the client}, including additions that add all files from $\overline{\mathsf{D}}$ into $\mathsf{EDB}$ and deletions that remove some of the existing entries from $\mathsf{EDB}$. Here, we follow a slightly different assumption compared to existing works~\cite{BlackstoneKM19,Gui21PP23}: the attacker knows $\overline{\mathsf{D}}$ are injected files, but not whether they are deleted in later operations. As mentioned, in $\mathsf{VIA}$, an attacker is assumed to be able to monitor the transcripts between the client and the server. Here we take the constructions having Type-I backward privacy as an example to illustrate our attacks.

\noindent{\bf Attack Description.} Let $(q, \tilde{q})$ be a pair of linked queries mentioned above, let $\mathsf{L}[w]=|\overline{\mathsf{D}}(w)|$ be a map that records the number of files containing $w$ and $\mathsf{V}[w]=\{\#f\}_{f \in \overline{\mathsf{D}}[w]}$ be multi-maps storing the size of the file containing $w$. Here both $\mathsf{L}$ and $\mathsf{V}$ are known to the attacker. Then the attacker can run Algorithm~\ref{Alg:baseline attack} to recover $q$.

The first step is to extract the insert volume of the above two queries (\lei{lines 1-3}). As all BP-DSSE constructions leak the update length pattern and response length pattern, for each pair of $q_i, \tilde{q}_i$, an attacker can learn their insert lengths
\begin{equation*}
a_{q}=(u_{q}+n_{q})/2~{\rm and}~a_{\tilde{q}}=(u_{\tilde{q}}+n_{\tilde{q}})/2
\end{equation*}
where $u_{q}=\mathsf{ulen}(\mathsf{EDB}, q)$ and $u_{\tilde{q}}=\mathsf{ulen}(\mathsf{EDB}, \tilde{q})$ are update lengths on $q$ and $\tilde{q}$ and  $n_{q}=\mathsf{rlen}(\mathsf{EDB}, q)$ and $n_{\tilde{q}_i}=\mathsf{rlen}(\mathsf{EDB}, \tilde{q})$ are the response lengths. With the insert length, the attacker can calculate the insert length $a_{\tilde{q}}-a_{q}$ of the keyword between $q$ and $\tilde{q}$ and use it to initialize the set of candidates $C_{q}$ for ${q}$ (\lei{line 4}). Specifically, it defines
\begin{equation*}
C_{q} = \{w: a_{\tilde{q}}-a_{q} = \mathsf{L}[w]\}.
\end{equation*}
Now, if there is only one keyword in the candidate set, the attacker directly returns it as the output of the query (\lei{lines 5-6}). Otherwise, the attacker further proceeds with the candidates by comparing the file volumes, removing the keyword whose corresponding file volumes do not include the response file volumes from the candidate set (\lei{lines 7-11}). More precisely, it narrows down the candidates to
\begin{equation*}
C_{q} = C_{q}/\{w: \exists~v \in \mathsf{fvol}[\mathsf{EDB}, \tilde{q}]/\mathsf{fvol}[\mathsf{EDB}, q], {\rm s.t.}~v \notin \mathsf{V}[w]\}
\end{equation*} 
where $\mathsf{fvol}[\mathsf{EDB}, q]$ and $\mathsf{fvol}[\mathsf{EDB}, \tilde{q}]$ are collections of file volumes returned by $q$ and $\tilde{q}$. If only one candidate remains, the attacker maps it to $q$. Otherwise, return $\bot$ (\lei{line 13}).

\begin{algorithm}[t]\small
\DontPrintSemicolon 
  \caption{\textbf{Toy Volumetric Inference Attack}}\label{Alg:baseline attack}
  \KwInput{Encrypted database $\mathsf{EDB}$, two linked queries $(q, \tilde{q})$, a multi- map $\mathsf{V} = \{(w, \{\#f\}_{f \in \overline{\mathsf{D}}[w]})\}$, where $\overline{\mathsf{D}}$ is the file collection inserted between $q$ and $\tilde{q}$, a map $\mathsf{L} = \{(w, |\overline{\mathsf{D}}[w]|)\}$}
  \KwOutput{The goal is to recover ${\bm q}$.}
    Count $u_{q} \leftarrow \mathsf{ulen}(\mathsf{EDB}, q)$ and $n_{q} \leftarrow \mathsf{rlen}(\mathsf{EDB}, q)$\;
    Count $u_{\tilde{q}} \leftarrow \mathsf{ulen}(\mathsf{EDB}, \tilde{q})$ and $n_{\tilde{q}} \leftarrow \mathsf{rlen}(\mathsf{EDB}, \tilde{q})$\;
    Compute $a_{q}=(u_{q}+n_{q})/2~{\rm and}~a_{\tilde{q}}=(u_{\tilde{q}}+n_{\tilde{q}})/2$ \;
    Set $\mathsf{C}_q \leftarrow \{w: \mathsf{L}[w] = a_{\tilde{q}}-a_{q}\}$ as the candidates\;
    \eIf{$\#(\rm{\mathsf{C}}_q)= 1$ }{
    Map $q$ to $w \in \mathsf{C}_q$ 
    }{
    Initialize a collection ${\bm v} \leftarrow \mathsf{fvol}[\mathsf{EDB}, \tilde{q}]/\mathsf{fvol}[\mathsf{EDB}, q]$\;
    \For{{\rm each} $w \in \mathsf{C}_q$}{
      remove $w$ from $\mathsf{C}_q$ if $\exists~v \in {\bm v}$ s.t. $v \notin \mathsf{V}[w]$ 
    }
    }
    Map $q$ to $w \in \mathsf{C}_q$ if \#$\rm{\mathsf{C}}_q$ = 1. Otherwise, return $\bot$
\end{algorithm}

\subsection{Attacks with Limited Prior Knowledge}\label{subsec:limited attack}
\noindent $\mathsf{VIA}$ describes a baseline attack to recover the query from insert volume patterns and file volume patterns under the standard forward and backward notions. Generally, the assumption that all added files are known to the adversary is strong. Besides, if the cleartext is not a public dataset, an attacker might not be able to get the file volumes. For these constraints, following we limit the basic assumption step by step, making our attack more practical.

\noindent{$\mathsf{PVIA}$: \bf{Attacking with Partial Inserted Files.}} Firstly, we consider the \emph{partially file known case} that only a fraction of files $\overline{\mathsf{D}}^{\dagger} \subset \overline{\mathsf{D}}$ are known to the adversary as most of existing works~\cite{BlackstoneKM19,NingHPYL0D21}. Then the attack inputs become to $\mathsf{V}[w] = \{\#f\}_{f \in \overline{\mathsf{D}}^{\dagger}[w]}$ and $\mathsf{L}[w] = |\overline{\mathsf{D}}^{\dagger}[w]|$. With reference to recommendations in~\cite{BlackstoneKM19}, we assume that known files are sampled uniformly from the entire file collection. The known file rate is parameterized by a value $\alpha \in \{0,1\}$. Under this setting, attackers with a fraction of files cannot perform equality matching with revealed leakage because the information derived from the known files will be inconsistent with that of the query response. To solve this problem, we modify the strategy (\lei{line 4}) to build the initial candidate set for query $q$ as
\begin{equation}\label{equ:candi}
{C}_q := \{w: \alpha a_{q} \le \mathsf{L}[w] \le a_{q}\}
\end{equation}
and choose the keyword
\begin{equation*}
\arg \max_{w \in {C}_q} |(\mathsf{fvol}[\mathsf{EDB}, \tilde{q}]/\mathsf{fvol}[\mathsf{EDB}, q]) \Cap \mathsf{V}[w]|
\end{equation*} 
as the attack result (\lei{line 13}).

\begin{figure}[t]
\begin{tikzpicture}
  \begin{axis}[
  enlarge x limits=0.02, 
enlarge y limits=0.003,ylabel=Response length growth, ymin=15,ymax=900,
symbolic x coords={$w_{0}$,$w_{1}$,$w_{2}$,$w_{3}$,$w_{4}$,$w_{5}$,$w_{6}$,$w_{7}$,$w_{8}$,$w_{9}$,$w_{10}$,$w_{11}$,$w_{12}$,$w_{13}$,$w_{14}$,$w_{15}$,$w_{16}$,$w_{17}$,$w_{18}$,$w_{19}$,$w_{20}$,$w_{21}$,$w_{22}$,$w_{23}$,$w_{24}$,$w_{25}$,$w_{26}$,$w_{27}$,$w_{28}$,$w_{29}$,$w_{30}$,$w_{31}$,$w_{32}$,$w_{33}$,$w_{34}$,$w_{35}$,$w_{36}$,$w_{37}$,$w_{38}$,$w_{39}$,$w_{40}$,$w_{41}$,$w_{42}$,$w_{43}$,$w_{44}$,$w_{45}$,$w_{46}$,$w_{47}$,$w_{48}$,$w_{49}$},
x tick label style={
rotate=90,
font=\footnotesize
},
legend style={nodes={scale=0.7}},
label style={font=\footnotesize},
y tick label style={
rotate=0,
font=\small
},
xtick distance=1,
width=9cm,
height=5cm,
]
\addplot[RYB1,mark=x] table[x=xdata,y=ydata]{
xdata ydata
$w_{0}$  405
$w_{1}$  296
$w_{2}$  343
$w_{3}$  302
$w_{4}$  335
$w_{5}$  298
$w_{6}$  251
$w_{7}$  219
$w_{8}$  169
$w_{9}$  131
$w_{10}$  193
$w_{11}$  184
$w_{12}$  146
$w_{13}$  161
$w_{14}$  142
$w_{15}$  143
$w_{16}$  126
$w_{17}$  127
$w_{18}$  111
$w_{19}$  95
$w_{20}$  106
$w_{21}$  77
$w_{22}$  131
$w_{23}$  107
$w_{24}$  100
$w_{25}$  34
$w_{26}$  97
$w_{27}$  124
$w_{28}$  87
$w_{29}$  94
$w_{30}$  76
$w_{31}$  93
$w_{32}$  88
$w_{33}$  139
$w_{34}$  84
$w_{35}$  48
$w_{36}$  62
$w_{37}$  79
$w_{38}$  81
$w_{39}$  78
$w_{40}$  57
$w_{41}$  100
$w_{42}$  76
$w_{43}$  80
$w_{44}$  89
$w_{45}$  70
$w_{46}$  66
$w_{47}$  138
$w_{48}$  75
$w_{49}$  67
};
\addlegendentry{Jan};
\addplot[RYB7,mark=+] table[mark=square*,x=xdata,y=ydata]{
xdata ydata
$w_{0}$  561
$w_{1}$  442
$w_{2}$  466
$w_{3}$  363
$w_{4}$  459
$w_{5}$  364
$w_{6}$  328
$w_{7}$  324
$w_{8}$  237
$w_{9}$  199
$w_{10}$  240
$w_{11}$  258
$w_{12}$  223
$w_{13}$  196
$w_{14}$  187
$w_{15}$  141
$w_{16}$  180
$w_{17}$  176
$w_{18}$  196
$w_{19}$  139
$w_{20}$  148
$w_{21}$  147
$w_{22}$  151
$w_{23}$  138
$w_{24}$  129
$w_{25}$  24
$w_{26}$  143
$w_{27}$  160
$w_{28}$  79
$w_{29}$  153
$w_{30}$  117
$w_{31}$  141
$w_{32}$  134
$w_{33}$  122
$w_{34}$  123
$w_{35}$  88
$w_{36}$  104
$w_{37}$  117
$w_{38}$  104
$w_{39}$  124
$w_{40}$  107
$w_{41}$  104
$w_{42}$  104
$w_{43}$  90
$w_{44}$  114
$w_{45}$  117
$w_{46}$  95
$w_{47}$  147
$w_{48}$  105
$w_{49}$  93
};
\addlegendentry{Feb};
\addplot[RYB3,mark=*] table[mark=square*,x=xdata,y=ydata]{
xdata ydata
$w_{0}$  524
$w_{1}$  432
$w_{2}$  426
$w_{3}$  340
$w_{4}$  422
$w_{5}$  350
$w_{6}$  311
$w_{7}$  282
$w_{8}$  287
$w_{9}$  177
$w_{10}$  241
$w_{11}$  231
$w_{12}$  189
$w_{13}$  196
$w_{14}$  177
$w_{15}$  173
$w_{16}$  167
$w_{17}$  157
$w_{18}$  170
$w_{19}$  135
$w_{20}$  161
$w_{21}$  176
$w_{22}$  160
$w_{23}$  134
$w_{24}$  129
$w_{25}$  14
$w_{26}$  124
$w_{27}$  128
$w_{28}$  100
$w_{29}$  115
$w_{30}$  117
$w_{31}$  144
$w_{32}$  116
$w_{33}$  123
$w_{34}$  92
$w_{35}$  65
$w_{36}$  99
$w_{37}$  105
$w_{38}$  119
$w_{39}$  121
$w_{40}$  101
$w_{41}$  134
$w_{42}$  90
$w_{43}$  107
$w_{44}$  106
$w_{45}$  115
$w_{46}$  94
$w_{47}$  190
$w_{48}$  108
$w_{49}$  105
};
\addlegendentry{Mar};
\addplot[RYB4,mark=star] table[mark=square*,x=xdata,y=ydata]{
xdata ydata
$w_{0}$  475
$w_{1}$  392
$w_{2}$  359
$w_{3}$  336
$w_{4}$  356
$w_{5}$  334
$w_{6}$  333
$w_{7}$  287
$w_{8}$  176
$w_{9}$  176
$w_{10}$  251
$w_{11}$  249
$w_{12}$  208
$w_{13}$  218
$w_{14}$  179
$w_{15}$  157
$w_{16}$  177
$w_{17}$  188
$w_{18}$  153
$w_{19}$  135
$w_{20}$  159
$w_{21}$  263
$w_{22}$  166
$w_{23}$  151
$w_{24}$  131
$w_{25}$  37
$w_{26}$  156
$w_{27}$  137
$w_{28}$  79
$w_{29}$  126
$w_{30}$  104
$w_{31}$  135
$w_{32}$  116
$w_{33}$  147
$w_{34}$  142
$w_{35}$  101
$w_{36}$  93
$w_{37}$  94
$w_{38}$  95
$w_{39}$  118
$w_{40}$  83
$w_{41}$  92
$w_{42}$  113
$w_{43}$  146
$w_{44}$  105
$w_{45}$  121
$w_{46}$  95
$w_{47}$  128
$w_{48}$  77
$w_{49}$  93
};
\addlegendentry{Apr};
\addplot[RYB5,mark=triangle*] table[mark=square*,x=xdata,y=ydata]{
xdata ydata
$w_{0}$  708
$w_{1}$  602
$w_{2}$  553
$w_{3}$  514
$w_{4}$  534
$w_{5}$  498
$w_{6}$  478
$w_{7}$  382
$w_{8}$  323
$w_{9}$  302
$w_{10}$  326
$w_{11}$  321
$w_{12}$  296
$w_{13}$  327
$w_{14}$  293
$w_{15}$  279
$w_{16}$  228
$w_{17}$  229
$w_{18}$  195
$w_{19}$  184
$w_{20}$  200
$w_{21}$  184
$w_{22}$  207
$w_{23}$  182
$w_{24}$  205
$w_{25}$  166
$w_{26}$  182
$w_{27}$  204
$w_{28}$  139
$w_{29}$  145
$w_{30}$  131
$w_{31}$  149
$w_{32}$  191
$w_{33}$  203
$w_{34}$  159
$w_{35}$  173
$w_{36}$  138
$w_{37}$  161
$w_{38}$  141
$w_{39}$  165
$w_{40}$  140
$w_{41}$  190
$w_{42}$  170
$w_{43}$  203
$w_{44}$  131
$w_{45}$  175
$w_{46}$  121
$w_{47}$  153
$w_{48}$  129
$w_{49}$  132
};
\addlegendentry{May};
\addplot[RYB6,mark=diamond*] table[mark=square*,x=xdata,y=ydata]{
xdata ydata
$w_{0}$  865
$w_{1}$  787
$w_{2}$  650
$w_{3}$  635
$w_{4}$  629
$w_{5}$  619
$w_{6}$  595
$w_{7}$  436
$w_{8}$  402
$w_{9}$  436
$w_{10}$  392
$w_{11}$  378
$w_{12}$  365
$w_{13}$  351
$w_{14}$  297
$w_{15}$  282
$w_{16}$  318
$w_{17}$  278
$w_{18}$  272
$w_{19}$  252
$w_{20}$  257
$w_{21}$  215
$w_{22}$  237
$w_{23}$  236
$w_{24}$  227
$w_{25}$  286
$w_{26}$  247
$w_{27}$  181
$w_{28}$  176
$w_{29}$  157
$w_{30}$  179
$w_{31}$  193
$w_{32}$  244
$w_{33}$  245
$w_{34}$  214
$w_{35}$  222
$w_{36}$  213
$w_{37}$  237
$w_{38}$  193
$w_{39}$  229
$w_{40}$  149
$w_{41}$  197
$w_{42}$  201
$w_{43}$  215
$w_{44}$  154
$w_{45}$  208
$w_{46}$  144
$w_{47}$  168
$w_{48}$  180
$w_{49}$  163
};
\addlegendentry{Jun};
\addplot[RYB1,thick] coordinates{($w_{25}$,80)($w_{24}$,380)};
\coordinate (spypoint) at (axis cs:$w_{28}$,180);
\coordinate (magnifyglass) at (axis cs:$w_{45}$,500);
\end{axis}
\draw[draw=RYB1,fill=none,thick]  (3.8,0.1) circle(0.2);
\node[pin=100:{%
    \begin{tikzpicture}[trim axis left,trim axis right]
    \begin{axis}[RYB1,
      symbolic x coords={$w_{24}$,$w_{25}$,$w_{26}$},
      ymin=10,ymax=40,
      x tick label style={font=\small},
      y tick label style={font=\small},
      line join=round,
      xticklabels = {$w_{24}$,$w_{25}$,$w_{26}$},
      enlargelimits,width = 3.5cm
    ]
    \addplot[RYB1,mark=x,thick] coordinates{($w_{24}$,100)($w_{25}$,34)($w_{26}$,97)};
    \addplot[RYB7,mark=+,thick] coordinates{($w_{24}$,129)($w_{25}$,24)($w_{26}$,143)};
    \addplot[RYB6,mark=*,thick] coordinates{($w_{24}$,129)($w_{25}$,14)($w_{26}$,124)};
    \addplot[RYB4,mark=star,thick] coordinates{($w_{24}$,131)($w_{25}$,37)($w_{26}$,156)};
    \end{axis}
    \end{tikzpicture}%
}] at (spypoint) {};
\end{tikzpicture}
\vspace{-20pt}
\caption{The growth of the number of files with respect to 50 randomly selected keywords in half a year}\label{fig:length-change}
\vspace{-10pt}
\end{figure}
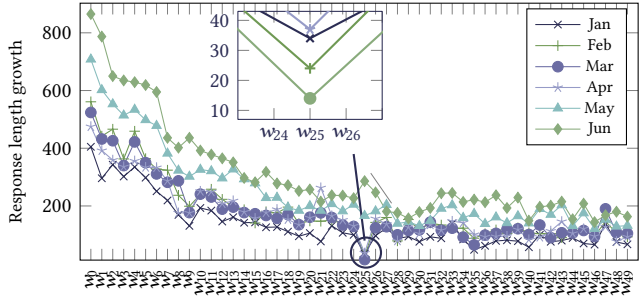

\noindent$\mathsf{LVIA}$: {\bf Attacking with Data Distribution Only.} To follow up on the above principle, we constrain the assumption further to the scenario where the attacker only captures the knowledge of data distribution which records the probability of a keyword appearing in a file. For ease of presentation, we directly parameterized this knowledge as the response length of each keyword in the dataset. Previous work~\cite{CashGPR15} has demonstrated that the harm of response length to query privacy is mild, owing to the fact that only a small proportion of queries hold a unique response length. However, the above result was obtained under static conditions. As shown in Figure~\ref{fig:length-change}, the update frequency of different keywords varies significantly across different time intervals in dynamic settings. 

Based on the above insight, we further optimize our attack by proceeding in rounds to reduce the number of candidates. Specifically, we can first obtain the candidate sets of the target ${C}_q^{(1)}, {C}_q^{(2)}, \cdots, {C}_q^{(n)}$ in different time intervals. Here, we choose the keyword candidates within a window of response length. The attack maps $q$ to the keyword that appears most frequently in the above candidates, i.e., 
\begin{equation*}
w= \arg \max\limits_{w \in \mathcal{W}} {\left\{|\mathcal{C}_w|: \mathcal{C}_w=\left\{C_q^{(i)}\right\}_{w \in C_q^{(i)}}\right\}},
\end{equation*}
where $\#\mathcal{C}_w$ is the number of candidate sets containing $w$.  Intuitively, large $\theta$ implies more elements in the candidate set and may require more rounds to select the correct candidate.


\subsection{Attacks without Prior Knowledge of Updates}\label{subsec:noupdate}
Previously, we have considered scenarios in which attackers possess prior knowledge of updates. Here we consider another scenario where they cannot touch such information \rev{but have access} to some historical knowledge of the encrypted database. Formally, we assume that the attacker possesses prior knowledge of the encrypted dataset before time $\mu_0$. This is a similar assumption as the one in LAAs against static SSE, where data breaches could happen previously or databases were not encrypted before. Following, we show that $\mathsf{IVA}$ can be adapted to this scenario after minor modification. 



Similarly, the first step is to find a linked query $\tilde{q}$ to the target query $q$, which occurs closest and after time $\mu_0$. In this case, the disclosed leakage from the query response of $\tilde{q}$ is closest to that derived from the attacker's prior knowledge, and the goal becomes recovering $\tilde{q}$. Following this principle, we use the approaches mentioned before to check the query equality between $q$ and previous queries. And if no such a $\tilde{q}$ can be found, we set $\tilde{q} = q$. Once the $\tilde{q}$  is determined, we initialize the candidates as 
\begin{equation*}
    C_{q} = C_{\tilde{q}} = \{w: a_{\tilde{q}} \ge \mathsf{L}[w]\}, 
\end{equation*}
where $a_{\tilde{q}}$ denotes the number of updates related to $\tilde{q}$ before $\tilde{q}$ happens. After that, we further narrow down the candidate set by leveraging the volume leakage and execute
\begin{equation*}
C_{q} = C_{q}/\{w: \exists~v \in \mathsf{fvol}[\mathsf{EDB}, \tilde{q}] {\rm s.t.}~v \notin \mathsf{V}[w]\}
\end{equation*}
Likewise, if only one candidate remains, the attacker outputs it as the attack result. The abovementioned approach can also be leveraged to tailor the $\mathsf{PVIA}$ and $\mathsf{LVIA}$ to suit the setting in this section. However, the attack performance of $\mathsf{LVIA}$ may be significantly weakened because the diversity of data distribution in updates cannot be used.

\section{Experiment Evaluations}\label{sec:experiment}
\noindent To validate our theoretical findings, we implement and evaluate the effectiveness of the proposed attacks under different attack settings. 
\subsection{Experiment Setup}
\noindent As with prior work, we focus on measuring attack accuracy by the proportion of recovered queries. All experiments are conducted on a server running Ubuntu using Intel Xeon 5218R CPU@2.10GHZ with 188G RAM\footnote{The code is publicly available at \url{https://github.com/FB-Attack/FB-Attack}}.

{\bf Experiment datasets.} The test datasets are extracted from two commonly-used email systems, Enron~\cite{Enron} and Lucene~\cite{Lucene}. We select 500, 1000, 2000, and 3000 most frequent words outside of the stemming as the keyword space and use these keywords and associated identifiers to build the encrypted database. For ease of presentation, we denote the encrypted database as $\mathsf{En}_i$ and $\mathsf{Lu}_i$ for Enron and Lucene datasets, respectively, where $i$ refers to the number of selected keywords.

{\bf Update operation.} Our experiments focus on two fundamental update operations: file addition and deletion. \rev{For additions, we simulate the dynamicity of the encrypted datasets by using emails from different time intervals. 
Specifically, if a query is generated
in July 2012, the query response only includes the items that are
inserted before that time. 
For deletions, we randomly delete existing emails with a predefined deletion ratio, where each file has the same probability of being deleted.}


\pgfplotstableread{
name
$\mathsf{En}_{1000}$
$\mathsf{Lu}_{1000}$
$\mathsf{En}_{1000}$
$\mathsf{Lu}_{1000}$
$\mathsf{En}_{1000}$
$\mathsf{Lu}_{1000}$
$\mathsf{En}_{1000}$
$\mathsf{Lu}_{1000}$
$\mathsf{En}_{1000}$
$\mathsf{Lu}_{1000}$
}\itemtable

\pgfplotstableread{
name
$\mathsf{En}_{2000}$
$\mathsf{Lu}_{2000}$
$\mathsf{En}_{2000}$
$\mathsf{Lu}_{2000}$
$\mathsf{En}_{2000}$
$\mathsf{Lu}_{2000}$
$\mathsf{En}_{2000}$
$\mathsf{Lu}_{2000}$
$\mathsf{En}_{2000}$
$\mathsf{Lu}_{2000}$
}\jtemtable

\pgfplotstableread{
name
$\mathsf{En}_{3000}$
$\mathsf{Lu}_{3000}$
$\mathsf{En}_{3000}$
$\mathsf{Lu}_{3000}$
$\mathsf{En}_{3000}$
$\mathsf{Lu}_{3000}$
$\mathsf{En}_{3000}$
$\mathsf{Lu}_{3000}$
$\mathsf{En}_{3000}$
$\mathsf{Lu}_{3000}$
}\ktemtable

\pgfplotstableread{
Clusters mean1  mean2 
1        0.010  0.008
2        0.009  0.002
4        0.146  0.011  
5        0.160  0.004  
7        0.391  0.002  
8        0.396  0.008  
10       0.617  0.004  
11       0.624  0.012  
13       0.839  0.006  
14       0.845  0.018
}\rrtable

\pgfplotstableread{
Clusters mean1  mean2 
1        0.931  0.001  
2        0.930  0.002  
4        0.839  0.006  
5        0.845  0.019  
7        0.708  0.024  
8        0.631  0.130  
10       0.592  0.057  
11       0.493  0.213  
}\rtable

\pgfplotstableread{
Clusters mean1  mean2 
1        0.592  0.057  
2        0.493  0.213  
4        0.741  0.068  
5        0.582  0.249  
7        0.799  0.069  
8        0.609  0.275  
10       0.829  0.072  
11       0.631  0.285
}\lltable

\pgfplotstableread{
Clusters mean1  mean2 
1        0.709  0.024  
2        0.631  0.130  
4        0.678  0.064  
5        0.591  0.170  
7        0.667  0.065  
8        0.576  0.185  
10       0.652  0.079  
11       0.543  0.217
13       0.613  0.119  
14       0.529  0.232
}\hhtable

\pgfplotstableread{
Clusters mean1 mean2 mean3
1        0.603  0.296  0.052
2        0.553  0.303  0.062
4        0.375  0.379  0.108
5        0.343  0.341  0.107
7        0.194  0.338  0.151
8        0.185  0.288  0.127
10       0.125  0.273  0.153
11       0.122  0.226  0.125
}\htable

\pgfplotstableread{
Clusters time1  time2  time3  time4
1        0.485  0.270  0.096  0.046
2        0.437  0.266  0.101  0.056
4        0.239  0.287  0.151  0.074
5        0.227  0.255  0.127  0.080
7        0.104  0.195  0.137  0.098
8        0.105  0.173  0.117  0.080
10       0.061  0.136  0.114  0.087
11       0.064  0.120  0.095  0.072
}\itable

\pgfplotstableread{
Clusters Length Size Comb
1        0.584  0.186  0.161
2        0.512  0.260  0.158
4        0.316  0.289  0.234
5        0.312  0.205  0.107
7        0.138  0.203  0.358
8        0.124  0.157  0.351
10       0.079  0.140  0.373
11       0.067  0.122  0.305
}\jtable

\pgfplotstableread{
Clusters time1  time2  time3  time4
1        0.157  0.234  0.227  0.221
2        0.160  0.236  0.228  0.222
4        0.198  0.255  0.237  0.241
5        0.195  0.251  0.237  0.232
7        0.217  0.255  0.244  0.241
8        0.216  0.245  0.241  0.239
10       0.227  0.253  0.245  0.244
11       0.226  0.245  0.242  0.241
}\stable
\pgfplotstableread{
Clusters ped06  ped12  ped18  ped24  
1        0.157  0.235  0.225  0.222
2        0.161  0.235  0.228  0.222
4        0.157  0.234  0.226  0.222
5        0.160  0.235  0.223  0.223
7        0.156  0.233  0.225  0.221
8        0.159  0.233  0.228  0.208
10       0.151  0.233  0.294  0.142
11       0.157  0.227  0.226  0.208
13       0.146  0.230  0.217  0.201
14       0.156  0.233  0.208  0.207  
}\ttable

\pgfplotstableread{
name
$\mathsf{En}_{500}$
$\mathsf{Lu}_{500}$
$\mathsf{En}_{1000}$
$\mathsf{Lu}_{1000}$
$\mathsf{En}_{2000}$
$\mathsf{Lu}_{2000}$
$\mathsf{En}_{3000}$
$\mathsf{Lu}_{3000}$
}\datatable

\begin{figure*}[!ht]
\center
\subfigure[T1 and T2 with different $\tau$ ($\lambda=120{\rm K}$)]{
\begin{tikzpicture}[scale=0.69]
\begin{axis}[ybar stacked,
    legend style={ legend columns=-1,
    legend pos=north east
    },  
    xtick=data,
    bar width=2mm,
    ymin=0,
    width=0.48\textwidth,
    height=0.3\textwidth,
    xticklabels from table={\itemtable}{name},
    x tick label style={rotate=30,anchor=east},
    label style={font=\large},
    ymax=1.3,
    ylabel={Attack Accuracy},
    area legend]    
    \addplot [RYB5,fill=RYB5,x tick label style={xshift=-0.3cm}] table[x=Clusters,y=mean1] {\rrtable};
    \addlegendentry[]{T2};
    \addplot [RYB3,fill=RYB3,x tick label style={xshift=-0.3cm}] table[x=Clusters,y=mean2] {\rrtable};
    \addlegendentry{T1};
    \end{axis}
        \draw (0.00,-1.0)  node[font=\small]{$\tau$:};
        \draw (0.80,-1.0)  node[font=\small]{$1$};
        \draw (2.20,-1.0)  node[font=\small]{$6$};
        \draw (3.60,-1.0)  node[font=\small]{$12$};
        \draw (5.00,-1.0)  node[font=\small]{$18$};
        \draw (6.30,-1.0)  node[font=\small]{$24$};
\end{tikzpicture}
\label{fig:fre-1}
}
\subfigure[T1 and T2 with different datasets]{
\begin{tikzpicture}[scale=0.69]
\begin{axis}[ybar stacked,
    legend style={ legend columns=-1,
    legend pos=north east
    },  
    xtick=data,
    bar width=2mm,
    ymin=0,
    width=0.48\textwidth,
    height=0.3\textwidth,
    xticklabels from table={\datatable}{name},
    x tick label style={rotate=30,anchor=east},
    label style={font=\large},
    ymax=1.3,
    ylabel={Attack Accuracy},
    area legend]    
    \addplot [RYB5,fill=RYB5,x tick label style={xshift=-0.3cm}] table[x=Clusters,y=mean1] {\rtable};
    \addlegendentry[]{T2};
    \addplot [RYB3,fill=RYB3,x tick label style={xshift=-0.3cm}] table[x=Clusters,y=mean2] {\rtable};
    \addlegendentry{T1};
    \end{axis}
      \draw (3.2,-1.0)  node[font=\small]{$\lambda = {120{\rm K}}$, $\tau = 24$};
\end{tikzpicture}
\label{fig:fre-2}
}
\subfigure[T1 and T2 with different $\lambda$ ($\tau = 24$)]{
\begin{tikzpicture}[scale=0.69]
\begin{axis}[ybar stacked,
    legend style={ legend columns=-1,
    legend pos=north east
    },  
    xtick=data,
    bar width=2mm,
    ymin=0,
    width=0.48\textwidth,
    height=0.3\textwidth,
    xticklabels from table={\ktemtable}{name},
    x tick label style={rotate=30,anchor=east},
    label style={font=\large},
    ymax=1.3,
    ylabel={Attack Accuracy},
    area legend]    
    \addplot [RYB5,fill=RYB5,x tick label style={xshift=-0.3cm}] table[x=Clusters,y=mean1] {\lltable};
    \addlegendentry[]{T2};
    \addplot [RYB3,fill=RYB3,x tick label style={xshift=-0.3cm}] table[x=Clusters,y=mean2] {\lltable};
    \addlegendentry{T1};
    \end{axis}
        \draw (0.00,-1.0)  node[font=\small]{$\lambda$:};
        \draw (0.90,-1.0)  node[font=\small]{$120{\rm K}$};
        \draw (2.60,-1.0)  node[font=\small]{$240{\rm K}$};
        \draw (4.50,-1.0)  node[font=\small]{$360{\rm K}$};
        \draw (6.30,-1.0)  node[font=\small]{$480{\rm K}$};
\end{tikzpicture}
\label{fig:fre-3}
}
\subfigure[T1 and T2 with different $\gamma$ ($\tau =24$)]{
\begin{tikzpicture}[scale=0.69]
\begin{axis}[ybar stacked,
    legend style={ legend columns=-1,
    legend pos=north east
    },  
    xtick=data,
    bar width=2mm,
    ymin=0,
    width=0.48\textwidth,
    height=0.3\textwidth,
    xticklabels from table={\jtemtable}{name},
    x tick label style={rotate=30,anchor=east},
    label style={font=\large},
    ymax=1.3,
    ylabel={Attack Accuracy},
    area legend]    
    \addplot [RYB5,fill=RYB5,x tick label style={xshift=-0.3cm}] table[x=Clusters,y=mean1] {\hhtable};
    \addlegendentry{T2};
    \addplot [RYB3,fill=RYB3,x tick label style={xshift=-0.3cm}] table[x=Clusters,y=mean2] {\hhtable};
    \addlegendentry{T1~};
    \addlegendentry{combine};
    \end{axis}
        \draw (0.00,-1.0)  node[font=\small]{$\gamma$:};
        \draw (0.90,-1.0)  node[font=\small]{$0\%$};
        \draw (2.30,-1.0)  node[font=\small]{$5\%$};
        \draw (3.60,-1.0)  node[font=\small]{$10\%$};
        \draw (5.00,-1.0)  node[font=\small]{$15\%$};
        \draw (6.40,-1.0)  node[font=\small]{$20\%$};
\end{tikzpicture}
\label{fig:fre-4}
}
\subfigure[T2 with different $\tau$ and $\gamma$ (${\gamma = 12{\rm K}}$)]{
\begin{tikzpicture}[scale=0.69]
\begin{axis}[ybar stacked,
    legend style={ legend columns=2,
    legend pos=north east
    },  
    xtick=data,
    bar width=2mm,
    ymin=0,
    width=0.48\textwidth,
    height=0.3\textwidth,
    xticklabels from table={\itemtable}{name},
    x tick label style={rotate=30,anchor=east},
    label style={font=\large},
    ymax=1.3,
    ylabel={Attack Accuracy},
    area legend]    
    \addplot [RYB5,fill=RYB5,x tick label style={xshift=-0.3cm}] table[x=Clusters,y=ped06] {\ttable};
    \addlegendentry{$\tau=6$};
        \addplot [RYB3,fill=RYB3,x tick label style={xshift=-0.3cm}] table[x=Clusters,y=ped12] {\ttable};
    \addlegendentry{$\tau=12$};
    \addplot [RYB6,fill=RYB6,x tick label style={xshift=-0.3cm}] table[x=Clusters,y=ped18] {\ttable};
    \addlegendentry{$\tau=18$};
    \addplot [RYB7,fill=RYB7,x tick label style={xshift=-0.3cm}] table[x=Clusters,y=ped24] {\ttable};
    \addlegendentry{$\tau=24$};
    \end{axis}
        \draw (0.00,-1.0)  node[font=\small]{$\gamma$:};
        \draw (0.90,-1.0)  node[font=\small]{$0\%$};
        \draw (2.30,-1.0)  node[font=\small]{$5\%$};
        \draw (3.60,-1.0)  node[font=\small]{$10\%$};
        \draw (5.00,-1.0)  node[font=\small]{$15\%$};
        \draw (6.40,-1.0)  node[font=\small]{$20\%$};
\end{tikzpicture}
\label{fig:fre-5}
}
\subfigure[T2 with noised query distribution]{
\begin{tikzpicture}[scale=0.69]
\begin{axis}[ybar stacked,
    legend style={ legend columns=1,
    legend pos=north east
    },  
    xtick=data,
    bar width=2mm,
    ymin=0,
    width=0.48\textwidth,
    height=0.3\textwidth,
    xticklabels from table={\datatable}{name},
    x tick label style={rotate=30,anchor=east},
    label style={font=\large},
    ymax=1.3,
    ylabel={Attack Accuracy},
    area legend]    
    \addplot [RYB5,fill=RYB5,x tick label style={xshift=-0.3cm}] table[x=Clusters,y=Length] {\jtable};
    \addlegendentry{U(0,10)};
    \addplot [RYB3,fill=RYB3,x tick label style={xshift=-0.3cm}] table[x=Clusters,y=Size] {\jtable};
    \addlegendentry{U(0,5)~};
    \addplot [RYB6,fill=RYB6,x tick label style={xshift=-0.3cm}] table[x=Clusters,y=Comb] {\jtable};
    \addlegendentry{U(0,0)};
    \end{axis}
         \draw (3.1,-1.0)  node[font=\small]{$\lambda = 120{\rm K}, \gamma= 0\%, \tau =24$};
\end{tikzpicture}
\label{fig:fre-6}
}
\vspace{-5pt}
 \caption{Attack results on query frequency matching attack. The backward private schemes that leak the query equality pattern are denoted as T1, while those that do not are denoted as T2. Let $\gamma \in \{0\%, 5\%, 10\%, 15\%, 20\%\}$ be the deletion rates and $\lambda \in \{120{\rm K}, 240{\rm K}, 360{\rm K}, 480{\rm K}\}$ be the number of queries issued in the total of query time intervals. Let $\tau \in \{1, 6, 12, 18, 24\}$ be the number of time intervals of query distributions given to the attacker. U$[0,n]$ implies the query distribution the attacker learned is perturbed by noise uniformly sampled from $[0,n]$. \rev{We use U$[0,0]$ to denote the case where no noise is introduced.}}\label{fig:frequency-commparison1}
\end{figure*}
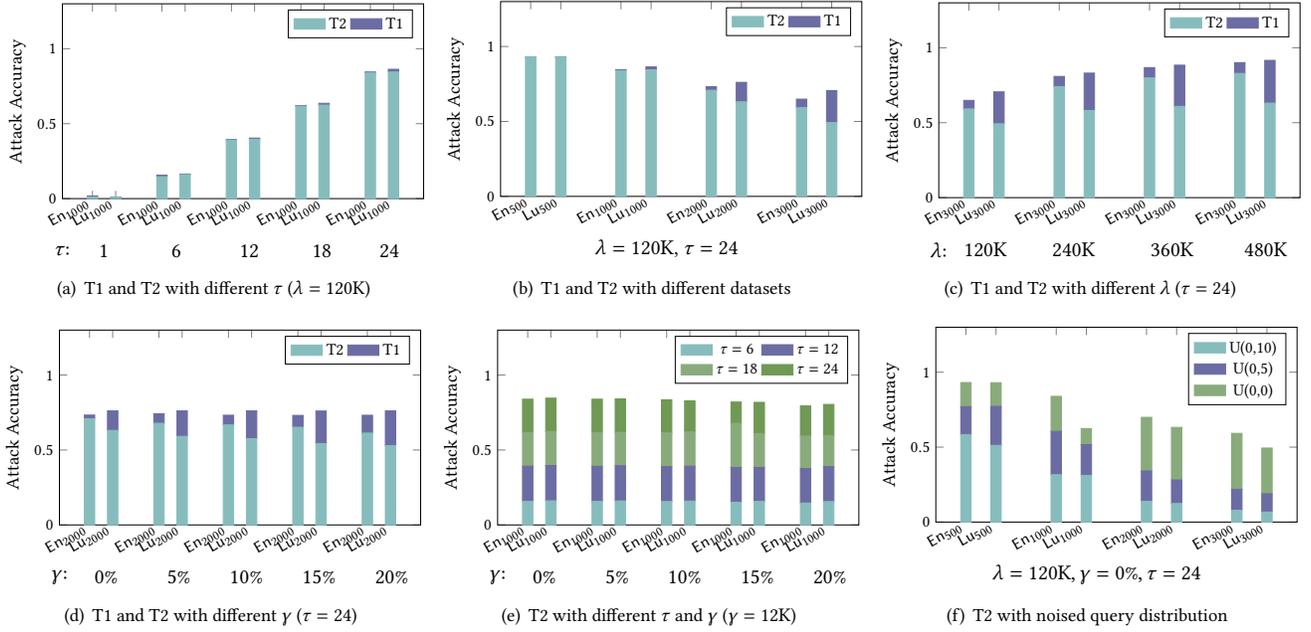

{\bf Auxiliary data.} Like most existing works, we assume that the passive \rev{attacker} learns some auxiliary data from the dataset. Here we consider the following three types of auxiliary data: 

\begin{itemize}
  \item {\bf Query distribution.} It records how frequently a keyword is queried in a period of time. We derive the query distribution from Google Trend~\cite{Googletrends}. For dynamicity concerns, query distribution here is defined monthly.
  \item {\bf Partial dataset.} A subset of files in the clear is assumed to be known by the attacker. This is the most commonly assumed information an attacker possesses in the literature on attacking an encrypted dataset. 
  \item {\bf Data distribution.} This statistical information characterizes the distribution of keywords across the entire dataset. Equivalently, it can be represented by the number of files related to each keyword. 
\end{itemize}

\subsection{Results on Frequency Matching Attacks}
\noindent We now present our evaluations on query frequency matching attacks introduced in Section~\ref{sec:FMA}, which exploits the query frequency information to recover the query. 
In this experiment, the query distributions on keywords are known to the attacker. The detailed settings are given below. 

Here, we consider two kinds of forward and backward private constructions, the schemes that leak the query equality pattern by relations of tokens are denoted as T1, while those that do not are denoted as T2. 
For the latter, the attacker must build the query equality by their response similarity. 
Without loss of generality, we set $\delta = 0.6$ here. 
To effectively model the dynamic nature of query distributions, we divide the entire query time into $\tau \in \{1, 6, 12, 18, \dots, 24\}$ intervals, wherein queries within each interval adhere to a distinct distribution. 
Similar to Oya et al.'s work~\cite{Simon21}, we sample the number of queries in each interval  from a Poisson distribution with parameter $\lambda/\tau$, where $\lambda\in \{120{\rm K}, 240{\rm K}, 360{\rm K}, 480{\rm K}\}$ is the expected total number of queries; 
we also assume that the number of queries for each keyword in each time interval follows a multinomial distribution, with the probabilities derived from the query frequency of the corresponding time interval of Google Trends~\cite{Googletrends}.
The deletion rate in the experiment is parameterized by $\gamma \in \{0\%, 5\%, 10\%, 15\%, 20\%\}$.

We start with a comparative experiment to show the advantage of the frequency matching attack, which exploits the diversity of query distribution in dynamic scenarios. By setting $\tau$ from $1$ to $24$, $\mathsf{FMA}$ gains a significant improvement in the recovery rate, as illustrated in Figure~\ref{fig:fre-1}. For schemes of T1 type, the recovery rate increases from 1.8\% to 84.5\%, showing an increase of 82.7\%. And for schemes of T2 type, we obtained a comparable outcome, thus highlighting the benefit of utilizing response similarity to link queries executed on identical keywords. Moreover, it is observed that the recovery rate for schemes of T2 type diminishes marginally as compared to that of T1. This is because rebuilding the query equality in T2 has a certain probability of being wrong.

\pgfplotstableread{
Clusters Length Size Comb
1        0.960  0.018  0.013
2        0.887  0.059  0.034
4        0.928  0.024  0.019
5        0.790  0.114  0.045
7        0.861  0.072  0.023
8        0.660  0.151  0.103
10       0.722  0.151  0.065
11       0.440  0.256  0.142
13       0.354  0.308  0.134
14       0.368  0.138  0.233
}\atable

\pgfplotstableread{
Clusters Length Size Comb
1        0.9625  0.013   0.0115
2        0.911   0.0375  0.025
4        0.9395  0.0185  0.0195
5        0.854   0.0630  0.032
7        0.892   0.046   0.020
8        0.7755  0.091   0.0585
10       0.802   0.0985  0.0395
11       0.6425  0.1465  0.0785
13       0.552   0.270   0.0875
14       0.417   0.263   0.141
}\btable

\pgfplotstableread{
Clusters Length Size Comb
1        0.9600  0.0106  0.0157
2        0.9057  0.0363  0.0250
4        0.9323  0.0223  0.0193
5        0.8607  0.0543  0.0273
7        0.8937  0.0410  0.0210
8        0.7997  0.0697  0.0049
10       0.8150  0.0853  0.0317
11       0.6987  0.1150  0.0647
13       0.552   0.270   0.0875
14       0.5170  0.2120  0.1023
}\ctable

\begin{figure*}[t]
\center
\subfigure[$\mathsf{PVIA}$ on $\mathsf{En}_{1000}$ and $\mathsf{Lu}_{1000}$]{
\begin{tikzpicture}[scale=0.69]
\begin{axis}[ybar stacked,
    legend style={ legend columns=-1,
    at = {(0.94,0.95)}
    },  
    xtick=data,
    bar width=2mm,
    ymin=0,
    width=0.48\textwidth,
    height=0.3\textwidth,
    xticklabels from table={\itemtable}{name},
    x tick label style={rotate=30,anchor=east}, 
    label style={font=\large},
    xlabel style={yshift=-5ex},
    ymax=1.3,
    ylabel={Attack Accuracy},
    area legend]    
    \addplot [RYB5,fill=RYB5,x tick label style={xshift=-0.3cm}] table[x=Clusters,y=Length] {\atable};
    \addlegendentry[]{$\gamma = 30\%$};
    \addplot [RYB3,fill=RYB3,x tick label style={xshift=-0.3cm}] table[x=Clusters,y=Size] {\atable};
    \addlegendentry{$\gamma = 20\%$};
    \addplot [RYB6,fill=RYB6,x tick label style={xshift=-0.3cm}] table[x=Clusters,y=Comb] {\atable};
    \addlegendentry{$\gamma = 10\%$};
    \end{axis}
        \draw (0.00,-1.0)  node[font=\small]{$\alpha$:};
        \draw (0.80,-1.0)  node[font=\small]{$90\%$};
        \draw (2.10,-1.0)  node[font=\small]{$80\%$};
        \draw (3.60,-1.0)  node[font=\small]{$70\%$};
        \draw (5.00,-1.0)  node[font=\small]{$60\%$};
        \draw (6.40,-1.0)  node[font=\small]{$50\%$};
\end{tikzpicture}
\label{fig:pkv-1000}
}
\subfigure[\textsf{PVIA} on \textsf{En}$_{2000}$ and \textsf{Lu}$_{2000}$]{
\begin{tikzpicture}[scale=0.69]
\begin{axis}[ybar stacked,
    legend style={ legend columns=-1,
    at = {(0.94,0.95)}
    },  
    xtick=data,
    bar width=2mm,
    ymin=0,
    width=0.48\textwidth,
    height=0.3\textwidth,
    xticklabels from table={\jtemtable}{name},
    x tick label style={rotate=30,anchor=east},
    label style={font=\large},
    xlabel style={yshift=-5ex},
    ymax=1.3,
    ylabel={Attack Accuracy},
    area legend]    
    \addplot [RYB5,fill=RYB5,x tick label style={xshift=-0.3cm}] table[x=Clusters,y=Length] {\btable};
    \addlegendentry[]{$\gamma = 30\%$};
    \addplot [RYB3,fill=RYB3,x tick label style={xshift=-0.3cm}] table[x=Clusters,y=Size] {\btable};
    \addlegendentry{$\gamma = 20\%$};
    \addplot [RYB6,fill=RYB6,x tick label style={xshift=-0.3cm}] table[x=Clusters,y=Comb] {\btable};
    \addlegendentry{$\gamma = 10\%$};
    \end{axis}
        \draw (0.00,-1.0)  node[font=\small]{$\alpha$:};
        \draw (0.90,-1.0)  node[font=\small]{$90\%$};
        \draw (2.30,-1.0)  node[font=\small]{$80\%$};
        \draw (3.60,-1.0)  node[font=\small]{$70\%$};
        \draw (5.00,-1.0)  node[font=\small]{$60\%$};
        \draw (6.40,-1.0)  node[font=\small]{$50\%$};
\end{tikzpicture}
\label{fig:pkv-2000}
}
\subfigure[$\mathsf{PVIA}$ on $\mathsf{En}_{3000}$ and $\mathsf{Lu}_{3000}$]{
\begin{tikzpicture}[scale=0.69]
\begin{axis}[ybar stacked,
    legend style={legend columns=-1,
    at = {(0.94,0.95)}
    },
    xtick=data,
    bar width=2mm,
    ymin=0,
    width=0.48\textwidth,
    height=0.3\textwidth,
    xticklabels from table={\ktemtable}{name},
    x tick label style={rotate=30,anchor=east},
    label style={font=\large},
    xlabel style={yshift=-5ex},
    ymax=1.3,
    ylabel={Attack Accuracy},
    area legend]    
    \addplot [RYB5,fill=RYB5,x tick label style={xshift=-0.3cm}] table[x=Clusters,y=Length] {\ctable};
    \addlegendentry[]{$\gamma = 30\%$};
    \addplot [RYB3,fill=RYB3,x tick label style={xshift=-0.3cm}] table[x=Clusters,y=Size] {\ctable};
    \addlegendentry{$\gamma = 20\%$};
    \addplot [RYB6,fill=RYB6,x tick label style={xshift=-0.3cm}] table[x=Clusters,y=Comb] {\ctable};
    \addlegendentry{$\gamma = 10\%$};
    \end{axis}
        \draw (0.00,-1.0)  node[font=\small]{$\alpha$:};
        \draw (0.90,-1.0)  node[font=\small]{$90\%$};
        \draw (2.30,-1.0)  node[font=\small]{$80\%$};
        \draw (3.60,-1.0)  node[font=\small]{$70\%$};
        \draw (5.00,-1.0)  node[font=\small]{$60\%$};
        \draw (6.40,-1.0)  node[font=\small]{$50\%$};
\end{tikzpicture}
\label{fig:pkv-3000}
}
\vspace{-5pt}
 \caption{Attack results of $\mathsf{PVIA}$ with different known file rates ($\alpha$) and deletion rates ($\gamma$), where $\alpha =90\%$ indicates that $10\%$ files are removed from the selected dataset and set as the attack's input and $\gamma = 30\%$ means the client deletes $30\%$ files from the encrypted dataset.}\label{fig:partial}
 \vspace{-5pt}
\end{figure*}
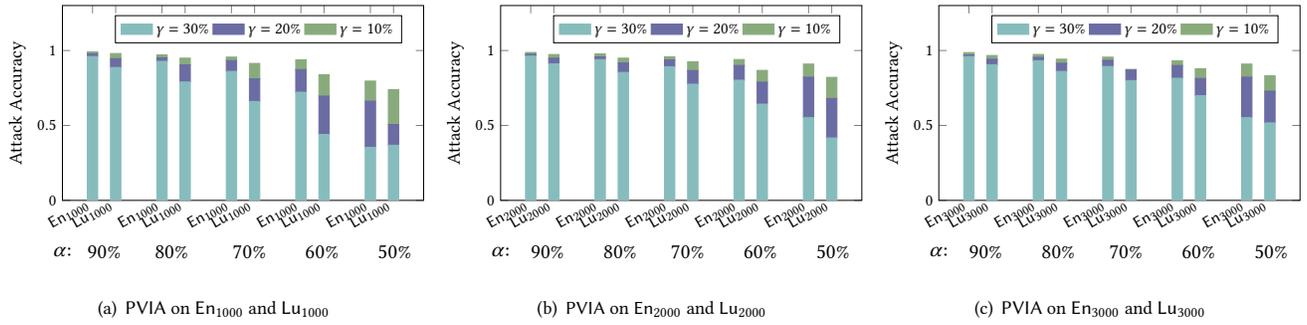

Further, we repeat this experiment on different datasets to verify the above results. The results in Figure~\ref{fig:fre-2} show that the recovery rate drops to 70.6\% for T1 and 49.3\% for T2 when attacking $\mathsf{Lu}_{3000}$. The main reason for this gap is that as the keyword space increases, the disparity in query frequency for each keyword diminishes, and in some cases, there is no difference at all, while the total number of queries remains unchanged. In addition, the average number of queries for each keyword is also reduced. As the number of queries is sampled, a small number of queries is insufficient to reflect the real-world query distribution. And the reason why the drop is more significant in T2 is that even for the identical keyword, their query responses can be significantly different in different stages due to frequent updates, which leads to incorrect query equality.

The above analysis shows that the number of observed queries may also influence attack performance. We continue examining the recovery rate under varying numbers of queries. Figure~\ref{fig:fre-3} lists the recovery rates of attacks given different numbers of queries. The recovery rate of T1 goes from 64.9\% to 90.1\% when the query volume grows from 120K to 480K. This result tells us that an attacker needs to collect as many queries as possible to launch the frequency matching attack, because only a large number of queries can get close to the true distribution due to the central limit theorem (CLT).

\pgfplotstableread{
Clusters Length Size 
1        0.075  0.697  
2        0.059  0.901  
4        0.042  0.526  
5        0.033  0.370  
7        0.022  0.313  
8        0.018  0.205  
10       0.017  0.217  
11       0.012  0.140  
}\xtable

\pgfplotstableread{
Clusters a1     a2      a3      a4      a5
1        0.084  0.071   0.054   0.049   0.051  
2        0.059  0.072   0.101   0.078   0.140  
4        0.049  0.046   0.037   0.041   0.049  
5        0.047  0.063   0.061   0.054   0.158  
7        0.011  0.017   0.016   0.024   0.076  
8        0.043  0.040   0.060   0.036   0.126  
10       0.012  0.012   0.021   0.011   0.056  
11       0.031  0.040   0.053   0.016   0.102  
}\ytable


\pgfplotstableread{
Clusters l1     l2      l3      l4      l5 
1        0.075  0.107   0.064   0.063   0.022  
2        0.059  0.077   0.074   0.243   0.177 
4        0.041  0.080   0.056   0.045   -0.005    
5        0.032  0.060   0.068   0.223   0.163
7        0.021  0.056   0.055   0.014   0.001 
8        0.018  0.039   0.051   0.197   0.142
10       0.016  0.039   0.053   0.004   0.001  
11       0.011  0.027   0.053   0.161   0.125 
}\ztable


\pgfplotstableread{
name
$\mathsf{En}_{500}$
$\mathsf{Lu}_{500}$
$\mathsf{En}_{1000}$
$\mathsf{Lu}_{1000}$
$\mathsf{En}_{2000}$
$\mathsf{Lu}_{2000}$
$\mathsf{En}_{3000}$
$\mathsf{Lu}_{3000}$
}\datatable

\begin{figure*}[t]
\center
\subfigure[\revs{Comparison of $\mathsf{LVIA}$-$\mathsf{S}$~{\rm and}~$\mathsf{LVIA}$-$\mathsf{K}$ ($\tau = 1, \eta =0.5$)}]{
\begin{tikzpicture}[scale=0.69]
\begin{axis}[ybar stacked,
    legend style={ legend columns=-1,
    legend pos=north east,
    },  
    xtick=data,
    bar width=2mm,
    ymin=0,
    width=0.48\textwidth,
    height=0.3\textwidth,
    xticklabels from table={\datatable}{name},
    x tick label style={rotate=30,anchor=east},
    label style={font=\large},
    ymax=1,
    ylabel={Attack Accuracy},
    area legend]    
    \addplot [RYB5,fill=RYB5,x tick label style={xshift=-0.3cm}] table[x=Clusters,y=Length] {\xtable};
    \addlegendentry[]{$\mathsf{LVIA}$-$\mathsf{S}$};
    \addplot [RYB3,fill=RYB3,x tick label style={xshift=-0.3cm}] table[x=Clusters,y=Size] {\xtable};
    \addlegendentry{$\mathsf{LVIA}$-$\mathsf{K}$};
    \end{axis}
\end{tikzpicture}\label{fig:length-1}
}
\hspace{-5pt}\subfigure[\revs{Comparison of $\mathsf{LVIA}$-$\mathsf{S}$ with different $\tau$ ($\eta = 0.5$)}]{
\begin{tikzpicture}[scale=0.69]
\begin{axis}[ybar stacked,
    legend style={ legend columns=4,
    legend pos=north east,
    },  
    xtick=data,
    bar width=2mm,
    ymin=0,
    width=0.48\textwidth,
    height=0.3\textwidth,
    xticklabels from table={\datatable}{name},
    x tick label style={rotate=30,anchor=east},
    label style={font=\large},
    ymax=1,
    ylabel={Attack Accuracy},
    area legend]    
    \addplot [RYB5,fill=RYB5,x tick label style={xshift=-0.3cm}] table[x=Clusters,y=l1] {\ztable};
    \addlegendentry[]{$\tau = 1$};
    \addplot [RYB3,fill=RYB3,x tick label style={xshift=-0.3cm}] table[x=Clusters,y=l2] {\ztable};
    \addlegendentry{$\tau = 3$};
    \addplot [RYB6,fill=RYB6,x tick label style={xshift=-0.3cm}] table[x=Clusters,y=l3] {\ztable};
    \addlegendentry{$\tau = 6$};
    \addplot [RYB7,fill=RYB7,x tick label style={xshift=-0.3cm}] table[x=Clusters,y=l4] {\ztable};
    \addlegendentry{$\tau = 18$};
    \end{axis}
\end{tikzpicture}\label{fig:length-2}
}
\hspace{-5pt}\subfigure[\revs{Comparison of $\mathsf{LVIA}$-$\mathsf{S}$ with different $\eta$ ($\tau =18$)}]{
\begin{tikzpicture}[scale=0.69]
\begin{axis}[ybar stacked,
    legend style={ legend columns=2,
    legend pos=north east,
    },  
    xtick=data,
    bar width=2mm,
    ymin=0,
    width=0.48\textwidth,
    height=0.3\textwidth,
    xticklabels from table={\datatable}{name},
    x tick label style={rotate=30,anchor=east},
    label style={font=\large},
    ymax=1,
    ylabel={Attack Accuracy},
    area legend]
    \addplot [RYB5,fill=RYB5,x tick label style={xshift=-0.3cm}] table[x=Clusters,y=a1] {\ytable};
    \addlegendentry[]{$\eta=0.1$};
    \addplot [RYB3,fill=RYB3,x tick label style={xshift=-0.3cm}] table[x=Clusters,y=a2] {\ytable};
    \addlegendentry{$\eta=0.2$};
    \addplot [RYB6,fill=RYB6,x tick label style={xshift=-0.3cm}] table[x=Clusters,y=a3] {\ytable};
    \addlegendentry{$\eta=0.3$};
     \addplot [RYB2,fill=RYB2,x tick label style={xshift=-0.3cm}] table[x=Clusters,y=a4] {\ytable};
    \addlegendentry{$\eta=0.4$};
     \addplot [RYB7,fill=RYB7,x tick label style={xshift=-0.3cm}] table[x=Clusters,y=a5] {\ytable};
    \addlegendentry{$\eta=0.5$};
    \end{axis}
\end{tikzpicture}
\label{fig:length-3}
}
\vspace{-10pt}
 \caption{Attack results of $\mathsf{LVIA}$ with different attack settings. $\mathsf{LVIA}$-$\mathsf{S}$ and $\mathsf{LVIA}$-$\mathsf{K}$ denote running $\mathsf{LVIA}$ is run in a sampled-data setting and known-data setting, respectively. $\tau$ denotes the number of groups (aka time intervals) that the dataset is partitioned, where $\tau=1$ denotes the static case. $\eta$ is the fraction of the dataset sampled to generate the data distribution.}\label{fig:commu-commparison1}
 \vspace{-5pt}
\end{figure*}
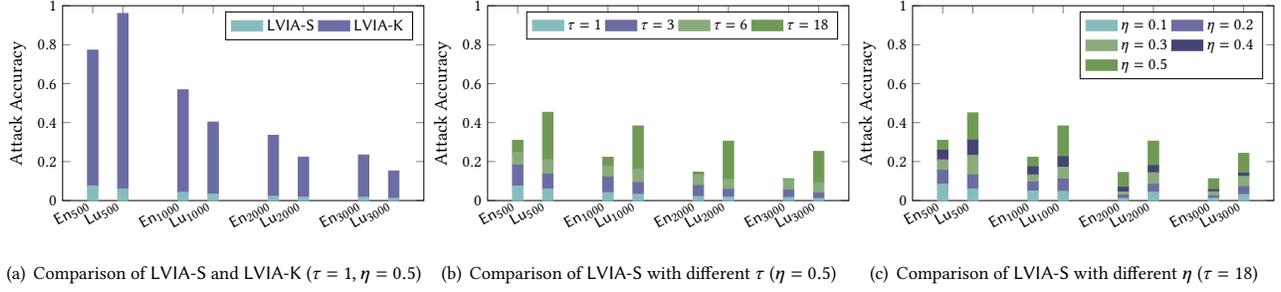

As data deletion is allowed in DSSE constructions, we also evaluate the ability of our attack to cope with this situation. We randomly remove a fraction of files from the dataset based on a specified rate, denoted as $\gamma$. The rest parameters are $\lambda =120{\rm K}$ and $\tau = 24$. As depicted in Figure~\ref{fig:fre-4}, the recovery rate of attacks on T1 schemes and $\mathsf{En}_{2000}$ remains stable at around 73.3\%, with little variation. This is because the T1 schemes can establish query equality through the relations of query tokens, which will not affect the group results. While for T2 schemes, the recovery rate decreased from 70.9\% to 61.3\% with the delete rate increases. The main reason is that the delete operations further enlarge the difference of query response associated with the same keyword, leading to incorrect query equality. We conducted more experiments with various parameter settings to investigate the impact of data deletion. Figure~\ref{fig:fre-5} shows that quite a few queries can be recovered under these settings, suggesting that $\mathsf{FMA}$ remains highly effective in compromising query privacy even in the presence of data deletions.

In reality, the query distribution possessed by attackers may not be entirely accurate, which can also affect the effectiveness of the proposed attack. To evaluate the performance of our attack against this case, we perturb the query distribution by adding a noise that follows the uniform distribution U$(0, n)$ and take it as the attacker's input. Likewise, we set $\lambda =120{\rm K}$, $\tau = 24$ and $n \in \{5, 10\}$. In this setting, the average number of queries per keyword in a month will be 10 for $\mathsf{En}_{500}$ and less than 2 for $\mathsf{En}_{3000}$. It is comparable to or even less than the magnitude of the noise. Unlike the above experiments, we cannot perform frequency matching to identify the potential candidates. Keywords whose query frequency is around (may not equal) that of their observations (group size) should be put into the candidate set. Figure~\ref{fig:fre-6} illustrates the attack results on datasets with various sizes. The results show that a certain number of queries can still be recovered under this weaker assumption, reflecting the vulnerability of DSSE in dynamic settings. For example, for query distribution perturbed by noise from distribution U$[0,10]$, the recovery rate of queried on dataset $\mathsf{En}_{500}$ is $58.4\%$. 

\subsection{Results on Volumetric Inference Attacks}
\noindent In this section, we detail the valuation for several volumetric inference attacks elaborated in Section~\ref{sec:tiv} which are all based on volumetric leakages. We run $\mathsf{VIA}$s under different settings and collect the corresponding recovery rates.

\noindent{\bf{$\mathsf{PVIA}$ attack.}} Recall that a $\mathsf{PVIA}$ attacker is assumed to have a subset of files in the clear and exploits response length, update length, and file size patterns. 
Response length and file size patterns could be obtained analogously as described in~\cite{Gui21PP23}. 
As for the update length pattern, \rev{we assume that each keyword will be queried after insertion in each time interval.
This assumption allows for the convenient computation of the update length pattern in each time interval during our evaluations. 
Otherwise, an attacker could also leverage timestamp information to obtain the update length pattern alternatively. Namely, our assumption is only for simplicity in our evaluations and could be omitted in real-world scenarios. 
} Suppose $\alpha$ and $\gamma$ be the known file rate and the delete rate, respectively, then the accuracy of attacking different datasets with different file known rates and deletion rates are presented in Figure~\ref{fig:partial}.

As seen in Figure~\ref{fig:pkv-1000}, when fixing $\gamma = 30\%$, with $\alpha$ decreasing from $90\%$ to $30\%$, the recovery rate on $\mathsf{En}_{1000}$ drops from $96\%$ to $35.4\%$. On the other hand, when fixing $\alpha = 60\%$, as $\gamma$ increases from $10\%$ to $30\%$, the recovery rate reduces from $83.8\%$ to $44\%$. The results show that both the known file rate and deletion rate significantly impact accuracy. Notably, this impact is achieved by affecting the volumetric information, particularly the length type information, including insert length and response length. While for the leakage of file volume, the unique combinations of file volumes can still be used to connect the query and keyword related to the same information. For example, if a file with a volume of 5 is returned for a query but the keyword "urgent" never appears in files with such a volume, we can remove "urgent" from the candidate set of the query. It remains worthwhile for us to validate the argument further. More attacks on other datasets can be seen in Figure~\ref{fig:pkv-2000} and Figure~\ref{fig:pkv-3000}.

\noindent{$\mathsf{LVIA}$ {\bf attack.}} The attack of $\mathsf{LVIA}$ is similar to $\mathsf{VIA}$ except that the auxiliary information captured by the adversary pertains to the data distribution rather than partial files in the targeted dataset. Therefore, the only available information that can be utilized is the length of the query response. 
\rev{Regarding this, we use a similar dataset to generate auxiliary information in the following experiments. 
Specifically, we partition the Enron/Lucene dataset into $\tau$ groups based on the time intervals each document belongs to. Within each interval, we further divide the documents into two halves: half of the documents serve as the attack target (i.e., data to be updated by the client in each interval), while (a fraction of) the other half is used to generate auxiliary distributions for the adversary.
Here, we also parameterize the number of intervals as $\tau$ and set $\tau \in \{1, 2, 3, 6, 18\}$, where $\tau = 1$ denotes the static case for prior passive LAAs. 
It is clear that increasing $\tau$ implies more refined distribution changes available to the adversary\footnote{However, having an excessive number $\tau$ of groups implies fewer documents in the auxiliary dataset of each group, which results in more noise in the auxiliary dataset.}. 
Based on this setting, we run a series of comparative experiments to investigate the influence of different attack settings on attack performance.}

\rev{The first experiment studies performance differences of $\mathsf{LVIA}$ between the known-data and sampled-data settings, where the sampled-data setting ($\mathsf{LVIA}$-$\mathsf{S}$) follows the description above and the known-data setting  ($\mathsf{LVIA}$-$\mathsf{K}$) assumes the adversary possesses the response length information of each keyword in the database. 
The results shown in Figure~\ref{fig:length-1} indicate a notable degradation in the performance of the attack when using noised auxiliary distributions (i.e., sampled-data)  compared to the exact auxiliary distributions.
For instance, the recovery rate on $\mathsf{En}_{500}$ reduces from $77.2\%$ to $7.5\%$, which is consistent with our insight given in Section~\ref{subsec:limited attack} and related works~\cite{BlackstoneKM19}. Namely, the impact of response length on query privacy is relatively mild in real-world applications, especially when the background knowledge is not consistent with the derived leakage from the encrypted database.}  

\rev{The experimental findings above show that the length-type leakage is somewhat inadequate regarding attack effectiveness. 
Fortunately, this shortcoming can be effectively alleviated in dynamic scenarios. As mentioned, the data distributions in different time intervals are normally different, providing adversaries with more information to refine leakage patterns and enhance attack accuracy. 
Briefly, we can identify candidate sets for an encrypted query in different time intervals and then select the most probable one (i.e., the one with the highest frequency of occurrence) as the recovered plaintext keyword. 
\revs{In light of this observation, we repeat $\mathsf{LVIA}$-$\mathsf{S}$ attack with $\tau \in \{1, 2, 3, 6, 18\}$ for Enron dataset from December 1st, 1999 to February 27th, 2001, and Lucene dataset from June 1st, 2006 to November 30th.} As depicted in Figure~\ref{fig:length-2}, the recovery rate exhibits a notable increase from $5.9\%$ (corresponding to the static case, $\tau=1$) to $44.9\%$ (corresponding to the dynamic case, $\tau=18$) when conducting the attack on $\mathsf{Lu}_{500}$ as expected.}

\rev{Finally, to examine the impact of data distribution quality on the attack accuracy, we utilize a fraction of the dataset (whose size is parameterized by the proportion $\eta \in \{0.1, 0.2, 0.3, 0.4, 0.5\}$) as the auxiliary dataset to generate the data distribution. \revs{As illustrated in Figure~\ref{fig:length-3}, while maintaining a constant number of time intervals $\tau = 18$, the recovery rate of $\mathsf{LVIA}$-$\mathsf{S}$ diminishes from $44.9\%$ to $5.9\%$ as $\eta$ decreases from $0.5$ to $0.1$.} This decline in accuracy occurs because the discrepancy between the data distributions obtained from the targeted dataset and the auxiliary dataset widens with the reduction of $\eta$.}

\subsection{Effectiveness on Prior Countermeasures}
\rev{The above evaluations offer a comprehensive analysis of existing FP/BP-DSSEs to ascertain how to combine leakages across updates. This allows us to adapt existing LAAs targeting static settings and make them effective against FP/BP-DSSEs. 
At this point, one may wonder if the above attacks can withstand any countermeasures similar to other LAAs. 
Indeed, the attacks presented in our work serve as an illustrative example of how to conduct passive attacks in FP/BP settings. Consequently, existing attacks (e.g., SAP~\cite{Simon21}, IHOP~\cite{OyaK22}, etc.) could be adapted by leveraging our proposed leakage framework to implement LAAs in dynamic settings.
Therefore, the ability to withstand countermeasures in dynamic settings relies on whether the static LAAs, to be adapted to dynamic settings, can also withstand the same countermeasures.
If the countermeasure is exceedingly strong, such as the almost zero-leakage dynamic construction proposed by George et al.~\cite{GeorgeKM21}, capable of thwarting all existing attacks, our attack will also fail.
If the design protects only volume patterns~\cite{Amjad23}, then frequency-based attacks would definitely work. 
Generally, one can transform strong attacks in static settings (e.g., SAP and IHOP) via our leakage framework for dynamic settings, and those transformed attacks (targeting specific dynamic constructions) will still be effective if no countermeasures are adopted correspondingly. In our experiments, we have evaluated our attacks w.r.t various leakage patterns, which implies the (in)effectiveness of different countermeasures, such as volume-hiding dynamic constructions~\cite{Amjad23} and others. }

\section{Discussion on Countermeasures}\label{sec:countermeasure}
\noindent Most recently, researchers have been aware of the importance of suppressing leakages in both update and search processes and made efforts to investigate effective techniques to suppress aforementioned leakages~\cite{GeorgeKM21, Amjad23, 0019ZD00J22} in dynamic scenarios. 
Intuitively, one can use both Oblivious RAM (ORAM)-like techniques and padding approaches to solve the above problem. ORAM-like techniques can be employed to conceal the search pattern and access pattern, whereas padding approaches can be used to eliminate the volume pattern. Unfortunately, the above techniques incur heavy communication, computation, or storage cost. Specifically, ORAM introduces a polylogarithmic search overhead, while worst-case padding can result in worst-case linear search time or quadratic index size~\cite{DemertzisPPS20}.

The above challenge motivates researchers to seek new efficient solutions. In a recent study, \citet{GeorgeKM21} extend their concept of (almost) zero leakage security from static to dynamic scenarios and proposed three new dynamic structure encryption schemes that could provide asymptotically better performance than using black-box ORAM simulation. The first two proposed solutions prioritize efficiency over achieving zero-leakage or perfect correctness, whereas the last solution addresses all the aforementioned issues at the cost of increased complexity and computational overhead. Aside from the above work, \citet{Amjad23} focus on hiding volume pattern for forward and backward private constructions. Their work tolerates the query equality pattern to obtain further efficiency improvement, implying that their work is only suitable for applications that access each datum uniformly. 

For security concerns, a simple and straightforward approach to address the problem stated in our work is to apply~\citet{GeorgeKM21}'s third construction here. However, all the above schemes are based on encrypted multi-maps (EMM). In prior works for file retrieving, EMMs typically refer to the inverted index, which aids in storing identifiers (value) and the corresponding keyword (label)~\cite{ChaseK10}. Therefore, the above solution only hides the leakages from one component of the overall DSSE scheme, which is the encrypted search index, and similar leakage patterns that occurred in the file retrieval phase can still be exploited to recover encrypted queries. The same insight has also been shown by~\citet{Gui21PP23} in the context of static settings.

To solve this problem, one possible solution is to substitute the stored values with the encrypted files instead of the identifiers in construction~\cite{GeorgeKM21}. Note that in this approach, one also needs to do padding for files such that all share the same volume to hide file volume information. It is not hard to see that such an approach incurs at least $\mathcal{O}(|\mathcal{W}|\max_{{w \in \mathcal{W}}}{|\mathsf{DB}[w]|}\cdot\max_{f \in {\bf D}}{\#(f)})$ storage cost and $\mathcal{O}(\max_{w \in \mathcal{W}}{|\mathsf{DB}[w]|} \cdot \max_{ f \in {\bf D}}\#(f))$ communication cost per query. 
For instance, let's consider using the Enron dataset as an example to create a multi-map for a dataset consisting of 30,204 emails and 73,991 keywords. In this case, the total storage overhead would increase by about $10^7\times$. Regarding this matter, finding a feasible approach that deserves to be deployed may still be a long-term endeavor.

\section{Conclusion}
In this paper, we revisit existing forward and backward private DSSE schemes and systematically analyze their limitations with regard to leakage-abuse attacks. By refining leakages, specifically for volumetric information of updates and queries and query equality, it is demonstrated that DSSE schemes supporting forward and backward privacy, are still vulnerable to LAAs. The considerable recovery rates underscore the need for solutions that completely thwart query equality leakage and protect volumetric leakage during updates, queries, and file retrieval.  By connecting to known results of prior arts, we observe that the solutions that can fully resolve the privacy problem in DSSE encounter severe performance issues, while the efficient solutions are still susceptible to real-world attacks. We expect that this work will deepen the understanding of the real-world security strength of forward and backward private DSSE schemes and shed light on the future design space for safeguarding DSSE schemes.

\section*{Acknowledgements}
This work is supported in part by the National Natural Science Foundation of China under Grant 62202228, by the Natural Science Foundation of Jiangsu Province under Grant BK20210330, by the Fundamental Research Funds for the Central Universities 30923011023, by HK RGC under Grants CityU 11217620, 11218322, R6021-20F, R1012-21, RFS2122-1S04, C2004-21G, C1029-22G, and N\_CityU139/21, by the Australian Research Council (ARC) Discovery Project under Grant No.DP200103308. We extend our sincere gratitude to the anonymous reviewers, Dr. Huayi Duan, and Dr. Shifeng Sun, whose constructive feedback and expert insights significantly improved the quality and rigor of this paper. Their thoughtful comments and suggestions were invaluable in refining the content and presentation.

\balance
\bibliographystyle{ACM-Reference-Format}
\bibliography{reference}

\appendix

\section{Security Notion of DSSE}\label{appendix:sec-notion}
\noindent The security of DSSE defines what an adversary learns in a DSSE scheme. It can be formalized as the indistinguishable model between a real word game $\mathsf{SSE}_{\mathsf{Real}}$ and an ideal world game $\mathsf{SSE}_\mathsf{Ideal}$ with predefined leakage profiles. These leakage profiles are usually modeled as triple $\mathcal{L} = (\mathcal{L}_{\mathsf{Setp}}, \mathcal{L}_{\mathsf{Srch}}, \mathcal{L}_{\mathsf{Updt}})$ define what operations $\mathsf{Setup}, \mathsf{Search}, \mathsf{Update}$ leak to the adversary. A secure DSSE scheme requires that the adversary learns nothing about the issued query and the content of the encrypted database except the above-admitted leakage profiles. Formally,

\begin{itemize}
  \item in game $\mathsf{DSSE}_{\mathsf{Real}}$, the DSSE scheme is executed honestly. The adversary observes the real transcript of each operation and outputs a bit $b$.
  \item in game $\mathsf{DSSE}_\mathsf{Ideal}$, the adversary sees a simulated transcript in place of the real transcript of the protocol. The simulated transcript is generated by a PPT algorithm $S$, known as the simulator, that has access to the leakage functions. Specifically, on $\mathsf{Setup(DB)}$, $S$ returns a transcript from leakages $S(\mathcal{L}_{\mathsf{Setp}}(\mathsf{DB}))$; and likewise for the $\mathsf{Search}$ and $\mathsf{Update}$ calls. The adversary then outputs a bit $b$.
\end{itemize}

\begin{definition}~\label{def:senmantic sec}
A DSSE scheme is said to be $\mathcal{L}$-adaptively-secure, with respect to a leakage function $\mathcal{L}$, if for any polynomial-time adversary $\mathcal{A}$ submitting a polynomial number of queries $q$, there exists a PPT simulator $\mathcal{S}$ such that:
\begin{equation*}
\left|{\rm Pr}[{\rm \mathsf{DSSE}}_{\mathsf{Real}}(\lambda, q) = 1] - {\rm Pr}[\mathsf{DSSE}_{\mathsf{Ideal}}(\lambda, q) = 1]\right| = {\rm negl}(\lambda)
\end{equation*}
where ${\rm negl}(\lambda)$ is a negligible function.
\end{definition}

\section{Notions of Forward and Backward Privacy}\label{appendix:forward and backward}
Despite the above security notion built on indistinguishability, DSSE also requires forward and backward privacy due to the disclosed update pattern and access pattern. 

For forward privacy, it requires that the newly added file/keyword pairs cannot be linked by previous query tokens. This implies that previous query tokens cannot be used to search the above newly added entries. In existing constructions, such a notion is usually achieved by refreshing the (partial) query token. Below is the formal definition of forward privacy.

\begin{definition}[Forward Privacy]
A $\mathcal{L}$-adaptively secure DSSE scheme is forward private iff the leakage function satisfies
\begin{equation*}
\mathcal{L}^{\mathsf{Updt}}(\mathsf{op}, \mathsf{in}) = \mathcal{L}^{\prime}(\mathsf{op}, \{\mathsf{id}_i, \kappa_i\})
\end{equation*}
where the set $\{\mathsf{ind}_i, \kappa_i\}$ is the set of modified file $\mathsf{ind}_i$ paired with the number $\kappa_i$ of modified keywords.
\end{definition}

While for backward privacy, it focuses on the privacy of entries which are added and deleted later. For different security and efficiency trade-offs, backward privacy is classified into three levels. Following, we borrow notions from Bost et al.'s~\cite{BostMO17} work to present the formal backward notion concisely. 
The first and strongest backward private notion is called backward privacy with insertion pattern, it leaks the files currently matching $w$, when they were inserted, and the total number of updates on $w$. Formally,

\begin{definition}[Backward Privacy with insertion pattern (Type-\uppercase\expandafter{\romannumeral1})]~\label{def:ibp}
A $\mathcal{L}$-adaptively-secure DSSE scheme is insertion pattern revealing backward-private iff the search and update leakage functions $\mathcal{L}^{\mathsf{Srch}}$ ,$\mathcal{L}^{\mathsf{Updt}}$ can be written as:
\begin{center}
$\mathcal{L}^{\mathsf{Updt}}(\mathsf{op}, w, \mathsf{id})=\mathcal{L}^{\prime}(\mathsf{op})$\\
$\mathcal{L}^{\mathsf{Srch}}(w)= \mathcal{L}^{\prime\prime}(\mathsf{TimeDB}(w), u_w)$
\end{center}
where $\mathcal{L}^{\prime}$ and $\mathcal{L}^{\prime\prime}$ are stateless.
\end{definition}

The second level of the backward private notion is called backward privacy with update pattern. It leaks the files currently matching $w$, when they were inserted, and when all the updates on $w$ happened (but not their content). Formally,

\begin{definition}[Backward Privacy with update pattern (Type-\uppercase\expandafter{\romannumeral2})]~\label{def:ubp}
A $\mathcal{L}$-adaptively-secure SSE scheme is update pattern revealing backward-private iff the search and update leakage functions $\mathcal{L}^{\mathsf{Srch}}$, $\mathcal{L}^{\mathsf{Updt}}$ can be written as:
\begin{center}
$\mathcal{L}^{\mathsf{Updt}}(\mathsf{op}, w, \mathsf{id})=\mathcal{L}^{\prime}(\mathsf{op}, w)$\\
$\mathcal{L}^{\mathsf{Srch}}(w)= \mathcal{L}^{\prime\prime}(\mathsf{TimeDB}(w), \mathsf{Updates}(w))$
\end{center}
where $\mathcal{L}^{\prime}$ and $\mathcal{L}^{\prime\prime}$ are stateless.
\end{definition}

The third level of the backward private notion is called weak backward privacy. It leaks the file currently matching $w$, when they were injected, when all the updates on $w$ happened, and deletion update canceled which insertion update. Formally, 

\begin{definition}[Weak Backward Privacy (Type-\uppercase\expandafter{\romannumeral3})]~\label{def:wbp}
A $\mathcal{L}$-adaptively-secure SSE scheme is weak backward-private iff the search and update leakage functions $\mathcal{L}^{\mathsf{Srch}}$ ,$\mathcal{L}^{\mathsf{Updt}}$ can be written as:
\begin{align*}
\mathcal{L}^{\mathsf{Updt}}(\mathsf{op}, w, \mathsf{id})=\mathcal{L}^{\prime}(\mathsf{op}, w)\\
\mathcal{L}^{\mathsf{Srch}}(w)= \mathcal{L}^{\prime\prime}(\mathsf{TimeDB}(w), \mathsf{DelHist}(w))
\end{align*}
where $\mathcal{L}^{\prime}$ and $\mathcal{L}^{\prime\prime}$ are stateless.
\end{definition}
Essentially, Def.~\ref{def:wbp} is proposed for practical considerations, where the major computations can be executed at the server side.

\section{Diversity in Dynamic Datasets}
Compared to static cases, the significant difference in dynamic databases is that the data in databases are changed over time. This change brings diversities of background knowledge, which can be exploited to improve the attack performance. In this section, we show the diversities in dynamic databases and explain their benefits to more powerful attacks.

The first dynamic property is query distribution which is used in $\mathsf{FMA}$. As mentioned, affected by factors such as data updates, seasonality, trending topics, news events, day of the week, and month of the year~\cite{Googletrends111}. As seen in Figure~\ref{fig:period-fre}, it reports the query distribution of 10 randomly selected keywords in April and August 2022. The query distribution of these selected keywords changes significantly. The second dynamic property is the diverse data distribution or so-called response volume (length) which is used in $\mathsf{VIA}$. Figure~\ref{fig:length-change} illustrates the volume distribution of 10 randomly selected keywords in different months. The graph shows that the number of files corresponding to each keyword also varies significantly across different time intervals.

Now, we take the diverse data distribution to show how it can affect the attack performance. The benefits of the above diversity to attack performances are two folders. On the one hand, such diversity increases the likelihood of a unique leakage occurring. unlike static databases, where only a small proportion of keywords have unique response volumes, dynamic databases provide more opportunities to track unique volumes. 
On the other hand, combining this diversity around different time intervals can further optimize the reconstruction space. One can compute the intersection of these reconstruction spaces of the query across multiple rounds to further narrow down the candidates. For instance, after the first update, the candidate keywords for $q$ are $\{w_1, w_2\}$, and after the second update, the candidates are $\{w_2, w_3\}$. By combining these two results, we can get that the actual keyword queried is $w_2$.

\end{document}